\documentclass[sigconf,balance=false,authorversion=true]{acmart}
\usepackage{popets}

\usepackage{bytefield}
\usepackage[acronym,nohypertypes={acronym},shortcuts,nonumberlist,nolist]{glossaries} 
\newacronym{ip}{IP}{Internet Protocol}
\newacronym{tlv}{TLV}{Type-Length-Value}
\newacronym{tcp}{TCP}{Transmission Control Protocol}
\newacronym{udp}{UDP}{User Datagram Protocol}
\newacronym{vpn}{VPN}{Virtual Private Network}
\newacronym{api}{API}{application programming interface}
\newacronym{ui}{UI}{user interface}
\newacronym{sql}{SQL}{Structured Query Language}
\newacronym{tls}{TLS}{Transport Layer Security}
\newacronym{uuid}{UUID}{Universally Unique Identifier}

\newacronym{cbc}{CBC}{cipher block chaining}
\newacronym{oaep}{OAEP}{Optimal Asymmetric Encryption Padding}

\newacronym{l2cap}{L2CAP}{Logical Link Control and Adaptation Protocol}
\newacronym{ikev2}{IKEv2}{Internet Key Exchange Protocol}
\newacronym{esp}{ESP}{IP Encapsulating Security Payloads}
\newacronym{nrlp}{NRLP}{Network Relay Link Protocol}
\newacronym{nwsc}{NWSC}{Network Service Connectors}
\newacronym{xpc}{XPC}{Inter-Process Communication}
\newacronym{ldm}{LDM}{Link Director Message}

\newacronym{ecg}{ECG}{electrocardiogram}

\newacronym{AEAD}{AEAD}{authenticated encryption with associated data}
\newacronym{mitm}{MitM}{machine-in-the-middle}
\newacronym{gcm}{GCM}{Galois/Counter Mode}
\newacronym{ecc}{ecc}{elliptic curve cryptography}
\newacronym{DH}{DH}{Diffie-Hellmann}
\newacronym{OAEP}{OAEP}{Optimal Asymmetric Encryption Padding}

\usepackage[colorinlistoftodos,prependcaption]{todonotes} 
\presetkeys%
    {todonotes}%
    {inline,backgroundcolor=orange}{}

\usepackage{lipsum}
\usepackage{soul}

\usepackage{etoolbox}
\newtoggle{blinded}
\togglefalse{blinded}

\usepackage{tabularx}
\usepackage{multirow}
\usepackage{colortbl}

\definecolor{witchred}{RGB}{185,15,34}
\definecolor{witchlightred}{RGB}{245,139,151}
\definecolor{witchblue}{RGB}{15,67,185}

\setcopyright{none}
\copyrightyear{2025}

\acmYear{2025}
\acmVolume{}
\acmNumber{}
\acmDOI{}
\acmISBN{}
\acmConference{}
\def\b{\discretionary{\mbox{$\hookleftarrow$}}{}{}}

\newcommand*\ShiftLeft{\ll}
\newcommand*\ShiftRight{\gg}

\newif\ifsubmission{}
\submissiontrue{}

\usepackage{tcolorbox}
\newtcolorbox{mybox}{colback=witchlightred!10!white,colframe=witchred}

\ifsubmission{}
    \newcommand{\review}[1]{}
\else
    \newcommand{\review}[1]{\begin{mybox}#1\end{mybox}}
\fi

\newcommand{\shortlink}[1]{\href{https://#1}{\texttt{#1}}}

\begin{document}

\title[WatchWitch]{WatchWitch: Interoperability, Privacy, and Autonomy for the Apple Watch}


\iftoggle{blinded}{\author{Authors removed for double-blind review.}}{

\author{Nils Rollshausen}
\orcid{0000-0003-2445-8684}
\affiliation{%
  \institution{Secure Mobile Networking Lab}
  \city{Technical University of Darmstadt} 
  \state{} 
  \country{Germany} 
}
\email{nrollshausen@seemoo.de}

\author{Alexander Heinrich}
\orcid{0000-0002-1150-1922}
\affiliation{%
  \institution{Secure Mobile Networking Lab}
  \city{Technical University of Darmstadt} 
  \state{} 
  \country{Germany} 
}
\email{aheinrich@seemoo.de}

\author{Matthias Hollick}
\orcid{0000-0002-9163-5989}
\affiliation{%
  \institution{Secure Mobile Networking Lab}
  \city{Technical University of Darmstadt} 
  \state{} 
  \country{Germany} 
}
\email{mhollick@seemoo.de}

\author{Jiska Classen}
\orcid{0009-0006-4341-2808}
\affiliation{%
    \institution{Hasso Plattner Institute}
    \city{University of Potsdam} 
    \country{Germany}
}
\email{jiska.classen@hpi.de}

}


\begin{abstract}
Smartwatches such as the Apple Watch collect vast amounts of intimate health and fitness data as we wear them.
Users have little choice regarding how this data is processed: The Apple Watch can only be used with Apple's iPhones, using their software and their cloud services.
We are the first to publicly reverse-engineer the watch's wireless protocols, which led to discovering multiple security issues in Apple's proprietary implementation.
With \textit{WatchWitch}, our custom Android reimplementation, we break out of Apple's walled garden---demonstrating practical interoperability with enhanced privacy controls and data autonomy.
We thus pave the way for more consumer choice in the smartwatch ecosystem, offering users more control over their devices.
\end{abstract}

\keywords{Wearables, Apple Watch, Privacy, Reverse Engineering}

\maketitle

\section{Introduction}

Of all our devices, our smartwatches know us best: Always on our wrists, they collect intimate data even as we sleep. As users, however, we have little control over this data.

The dominating smartwatch vendors---including Apple~\cite[p.~39–41]{antitrustUsa}, Samsung~\cite{galaxyWatchInterop}, and Google~\cite{pixelWatch}---all rely on ecosystem lock-in effects: For example, users can only use their Apple Watch with an iPhone, and no third-party smartwatch can offer the same level of integration with an iPhone as Apple's watches. Apple Watch owners are forced to use their watches on Apple's terms---using their devices, their software, and their cloud services. This leaves users of many smartwatches, from Apple or other vendors, with little control over their devices and nebulous promises of privacy.

This vendor lock-in has not gone unnoticed by legislators: An antitrust motion in the US identifies the Apple Watch as part of Apple's alleged smartphone monopoly and demands more interoperability~\cite{antitrustUsa}. Similar efforts are being implemented in the EU with the Digital Markets Act~\cite{euDigitalMarkets}. Apple claims to have researched interoperability for three years, concluding that it was infeasible~\cite{watchInterop9to5mac}.
\newpage

In this context, we introduce \textit{WatchWitch}: Based on extensive reverse-engineering of the Apple Watch, we create an open-source Android app that allows users to use their Apple Watch on their own terms and with an Android phone. Unlike the closed-source iOS system, Android lets users control much larger parts of the software stack. Our proof-of-concept \textit{WatchWitch} app shows that interoperability is feasible in practice and demonstrates how all users, iOS and Android, can benefit from using custom software built with their needs in mind---gaining independence from vendors with a history of privacy violations~\cite{breakmi, seemooFitbitVuln, googleFitbitTechradar, getHealthLeak}.

Beyond reimplementing features provided by Apple, \textit{WatchWitch} gives users better privacy controls and full autonomy over their data: We allow users to make fine-grained decisions about the watch's network connections through a user-controlled firewall and keep all their health data securely on-device. At the same time, we give them full access to the data their watch collects beyond what is displayed in Apple's standard \gls{ui}.

In the process of reverse-engineering the Apple Watch, we analyze its underlying security architecture---including how sensitive health data is protected in transit from the watch to the phone.

In short, our paper makes the following contributions:

\begin{enumerate}
    \item We analyze and document several previously unknown and widely deployed protocols used by the Apple Watch, thus opening them up to further security research.
    \item We provide a tool suite for working with and analyzing the Apple Watch's proprietary protocols.
    \item We demonstrate real-world interoperability using our open-source \textit{WatchWitch} app, supporting several core smartwatch features and maintaining hardware-backed security.
    \item We show usable privacy enhancements within \textit{WatchWitch}, giving users more control of---and insight into---the intimate data their watch collects.
    \item We analyze the security of the Apple Watch's wireless communication and discuss several vulnerabilities. The corresponding mitigations improve security for all users.
\end{enumerate}

The source code of WatchWitch, with further tooling and a demo video, can be found at \shortlink{github.com/seemoo-lab/watchwitch}.

\subsection*{Responsible Disclosure}

We disclosed two vulnerabilities to Apple in November 2023. Apple acknowledged both vulnerabilities.
They decided not to fix the first vulnerability as they claim it cannot be exploited (\autoref{sec:security:forging_health_data}).
They released a fix for our second issue in March 2024 (\autoref{sec:security:ike})\iftoggle{blinded}{}{~\cite{watchOSpatchnotes}}.
We reported a third issue in May 2024 and are awaiting a fix.

\section{Background}

We introduce smartwatches, the health data they collect and current interoperability efforts. Furthermore, we give background information on IPSec as it plays an important role for the Apple Watch.

\subsection{Smartwatches}

Smartwatches are wearable devices worn in place of an analogue watch. They offer a variety of features in addition to telling the time, typically including health and fitness tracking, instant messaging, and integration with a connected smartphone.
Smartwatches that go beyond simple fitness trackers can allow users to install third-party apps. This market segment includes the Apple Watch and various watches running Google's \textit{Wear OS} operating system. In this class of devices, Apple alone holds a market share of 45\%~\cite{appleWatchMarketShareCounterpoint}.

Only a few existing smartwatches focus on open hard- and software~\cite{pineTime, zsWatch, openSmartwatch}. Only PineTime goes beyond a simple hobbyist project, and none of them can compete with the capabilities of flagship commercial watches. Furthermore, some commercial smartwatches running Wear OS can be modified to run the Asteroid operating system~\cite{asteroidOS}, emphasizing user control and privacy at the cost of feature support.

When it comes to health and fitness data collection, even budget models typically come with heart rate sensors and step counters. What distinguishes the more expensive flagship models are more advanced sensors~\cite{galaxyWatchSpecs, appleWatchUltraSpecs}. These additional hardware capabilities are paired with more software features to enable precise tracking and estimation of health metrics.

Current models of the Apple Watch, which are the focus of our work, are consistently at the top of the market regarding sensor capabilities and software features. The current Apple Watch Ultra 2, for example, comes with electrical and optical heart sensors, a pulse oximeter, a skin temperature sensor, a GPS receiver, a gyroscope, and an accelerometer. Equipped with this hardware, it offers \gls{ecg} readings, extensive heart rate monitoring, blood oxygen measurements, activity recognition, sleep phase tracking, ovulation time estimation, and more~\cite{appleWatchUltraSpecs, appleWatchOximeter}.

\subsection{Health Data}
\label{sec:background:healthdata}

Modern smartwatches collect a trove of highly intimate health data. With watches offering sleep tracking~\cite{watchSleepTracking}, users are encouraged never to take off their devices. As a result, an Apple Watch may have access to a long-term, near-uninterrupted stream of sensor readings---a level of medical surveillance reserved to clinical settings only a few years prior.

If the very same data \textit{were} collected in a clinical setting, strict regulations, such as HIPAA~\cite{hipaa} in the US, would likely apply to storing, using, and sharing of said data. However, since smartwatches are not marketed as medical devices, most regulations on collecting and using medical information do not apply, giving manufacturers significantly more leeway to store, share, and sell user data~\cite{hipaapotamus}.

Consider how sensitive the information collected by these devices is: As certain bodies become more and more politicized, so does the data quantifying them. Period tracking apps on user's smartphones have already been used to prosecute unlawful abortions in the UK~\cite{tortoiseAbortion}. Thus, the long-term skin temperature recordings and user-entered cycle tracking information present on a smartwatch can become a dangerous liability for the user. Similar concerns apply to other measurements that could be used to infer a wide range of physical and mental health conditions.

When users opt to track their health using a smartwatch, they have to trust the device manufacturer in regard to this highly private data. Many manufacturers, however, do not have great track records when it comes to privacy and security~\cite{breakmi, seemooFitbitVuln, googleFitbitTechradar, getHealthLeak}.

While Apple claims that all health data is exchanged using end-to-end encryption even if shared using iCloud and remains inaccessible while the device is locked~\cite{appleHealthPrivacyWhitepaper}, users still have to trust Apple to maintain this behavior over future updates. As an additional risk factor, the same reasoning applies to any third-party app that is granted access to \textit{HealthKit} data. Large-scale leaks of private HealthKit data via third parties have already occurred~\cite{getHealthLeak}.

Given a sufficiently authoritarian government, Apple could also be compelled to disclose some types of data collected by their watches to government entities by altering their software. The Google-owned fitness tracker manufacturer Fitbit has already stated that they would comply with government requests for user health data, including period tracking information~\cite{googleFitbitTechradar}.

For all these reasons, we stress the need for smartwatch infrastructure that places us as users in control of our most intimate data---free from third parties and with full autonomy over our bodies and the data traces they leave behind.

\subsection{Interoperability}

\review{Following reviewer D's suggestion, we provide additional context to Apple's statement on exploring Android interoperability (now cited as \cite{watchInterop9to5mac}), why these efforts were not pursued further and how they conflict with Apple's business interests.}

Manufacturers often market smartwatches as companion products for their latest smartphone. With the deep integration between a smartwatch and smartphone being a key selling point, vendors have little incentive to ensure interoperability with other devices.

The Apple Watch in particular is incredibly tightly coupled with Apple's iPhone and iCloud ecosystem~\cite{watchHandoff,watchCameraRemote}, using proprietary protocols that are unavailable to third parties. Third-party watches can, for example, receive push notifications from an iPhone but not reply to them~\cite[p.~40]{antitrustUsa}. And while iPhones work with third-party smartwatches to some degree, the Apple Watch cannot be used with non-Apple smartphones in any way.

Unlike Apple's smartwatches, Wear OS devices can technically be used with iPhones. Recent flagship models from both Samsung and Google, however, have abandoned iOS compatibility due to concerns about an imperfect user experience~\cite{galaxyWatchInterop, pixelWatch}. Samsung also requires a paired Galaxy phone to use some features of its latest watches~\cite{galaxyWatchSpecs}. This means that users of either platform are essentially locked in to their choice of smartphone after purchasing a smartwatch at significant expense---and vice versa.

This is also part of the US antitrust complaint brought against Apple in March 2024. The complaint poses that Apple is deliberately limiting the use of third-party watches on iOS and restricting Apple Watch interoperability to expand their monopoly in the US smartphone market~\cite[pp.~39--41]{antitrustUsa}. According to reporting by 9to5Mac~\cite{watchInterop9to5mac}, Apple stated that they had researched possible Android interoperability for three years but ultimately concluded it was not technically feasible.

However, in \autoref{sec:reimplementation}, we show that such interoperability between the Apple Watch and an Android smartphone is indeed possible: We create a working Android prototype that supports core functionality and allows developers to add support for further services. We found that there were multiple technical difficulties Apple may have encountered, which we had to solve with WatchWitch. In particular, Android's network stack does not expose essential functionality on non-rooted devices, and storing health-related data with hardware-backed encryption is only supported on recent Android devices~\cite{rene-android-platform-security}. Switching communication from Bluetooth to \mbox{Wi-Fi} might also incur higher energy costs, lowering the watch's battery runtime. Nonetheless, the obstacles that limit our prototype (see \autoref{sec:limitations}) are not insurmountable and could likely be eliminated by Apple.

Aside from technical feasibility, the reason for Apple not providing Android interoperability could also lie in their interest not to weaken the iPhone's market position. The claims made in the antitrust complaint, which cites Apple-internal documents, match this reasoning~\cite[pp.~40]{antitrustUsa}. If the antitrust motion were to succeed and result in Apple opening up their proprietary interfaces, this would certainly benefit a wide range of consumers: The Apple Watch is a remarkable piece of hardware, and presumably many Android users would use it with their phones if given the ability to do so. Similarly, iPhone users could choose from a much wider variety of smartwatches while enjoying the same deep, seamless integration with their phones that an Apple Watch provides.

\subsection{Apple Watch Features}
\label{sec:background:features}

As a modern flagship smartwatch, the Apple Watch is designed to integrate tightly with a paired iPhone. But even on its own, the watch can measure and collect health data from its sensors, track workouts, play audio content, and more. When connected to \mbox{Wi-Fi} or cellular, users can also check weather and stock market data or install new apps from the App Store.

To use the full extent of the watch's capabilities, however, it has to be connected to its paired iPhone at least periodically. While the Apple Watch continues to collect data when not connected to the phone and allows users to add new workouts and other data points, the views of this data offered on the watch are minimal. Users can, for example, only check their current heart rate, current noise levels, and a summary of their last night's sleep.

A full overview of the user's health metrics with long-term data can only be seen in the Health app on the connected iPhone, which receives a copy of all collected samples when it connects to the watch. An active connection to the paired phone is also required to receive most notifications (iMessage being the exception), synchronize calendar events, contacts, and photos, share the phone's cellular connection, or access its camera.

Some of the Apple Watch's features, particularly those related to health measurements, are unavailable in some regions for regulatory reasons. This includes the ability to record \glspl{ecg} and detect irregular heart rhythm and atrial fibrillation~\cite{watchFeatureAvailability}. The pulse oximeter found on recent watch models cannot be used in the US due to a patent dispute~\cite{appleWatchOximeter}. With WatchWitch, we can circumvent these restrictions and enable features regardless of geographical location.

\subsection{IPSec}

The Apple Watch uses IPSec to create a \gls{vpn} connection between the watch and the phone. IPSec defines protocols to establish secure tunnels between two devices.

With the \gls{ikev2}~\cite{rfcIKEv2} the two devices negotiate cryptographic parameters and perform authenticated key exchanges. During an \gls{ikev2} handshake, parties typically use long-term asymmetric keys to establish ephemeral symmetric shared secrets. These ephemeral secrets are then used to encrypt application traffic between the devices using \gls{esp}. 
\gls{ikev2} is not designed to carry application traffic but can perform a range of signaling tasks, communicating parameters, network addresses, link states, etc. A single \gls{ikev2} message typically consists of several \textit{payloads}, containing different kinds of information---for example supported algorithms, key material, or certificates. Implementors may also extend the protocol with their own private payloads.

\Gls{esp} is only a thin wrapper protocol applying the encryption and authentication mechanism negotiated by \gls{ikev2} to application traffic using the given ephemeral key material.

\section{Methodology}
\label{sec:methodology}

\review{Following feedback from reviewers B and D, we incorporate more details about our reverse-engineering methodology and specific challenges we faced.}

Our work is based on reverse-engineering the network protocol stack of the iPhone and the Apple Watch. We look at the communication between both devices from two angles: \textbf{dynamically} by observing live communication and \textbf{statically} by analyzing files and system binaries. Our general reverse-engineering approach follows related work on iOS protocols~\cite{alexContinuity, toothpickerPaper, lpm, magicpairing, aristoteles, airtag}. In addition to these methods, we develop our reimplementation jointly with our reverse engineering in a process of \textit{iterative reimplementation}.

Before performing reverse engineering, we must ensure system-level access to our devices. Since iOS does not allow rooting, we use public jailbreaks to gain root access to the operating system. 

\subsection{Dynamic Analysis}

\paragraph{System logs} We inspect the iPhone and Apple Watch system logs using the macOS Console app. These logs give a first look into which processes are used when both devices exchange data.

\paragraph{Frida} We use Frida~\cite{frida}, a dynamic reverse-engineering toolkit, which allows intercepting functions in running processes on iOS. Whenever a process calls a function that we deem interesting, Frida intercepts the function execution and lets us read and modify variables.
Its JavaScript \gls{api} allows automating analysis of system functionality, such as dispatching data to threads and verbose logging~\cite{fridascripts}. This is crucial for us in dissecting the watch's different encryption layers, as it lets us automatically extract private key material and perform decryption and custom message parsing on the fly. Without this runtime access to the device, we could not gain any visibility into the exchanged messages, as every session uses fresh ephemeral key material.

\paragraph{Network interfaces} We observe Bluetooth and \mbox{Wi-Fi} interfaces directly using developer tools such as \texttt{tcpdump}~\cite{tcpdump} and Apple's \textit{Bluetooth Packet Logger}~\cite{appleBluetoothPacketLogger}. We use these tools to build a corpus of observed messages that helps us understand both the static message structure as well as the dynamic protocol behavior, such as the semantics of sequence numbers or different data streams.

\subsection{Static Analysis}

\paragraph{Binary analysis} Internally, multiple iOS processes handle communication between the iPhone and the Apple Watch. Using disassemblers we can more closely inspect the behavior of relevant functions. This lets us uncover codepaths that are not taken during regular usage: Most notably deprecated or rarely used messages and the behavior of protocols in the presence of various errors.

\subsection{Iterative Reimplementation}

We develop our reimplementation of the protocol stack simultaneously with our reverse engineering efforts. By iteratively integrating discovered protocols and features, we create a lab setup where we have full control over the messages sent by our implementation. This allows us to trigger previously unreachable behavior on the watch and lets us access the watch's deeply nested protocols in a targeted manner. As many of the watch's features are only used in specific scenarios or for a short period during connection establishment, this greatly aids our overall reverse engineering effort.

\subsection{Research Framework}

Using Frida, we create powerful custom tooling that is able to capture, decrypt, and analyze watch traffic in real time---including full parsing of exchanged messages across all layers of the protocol stack. As we expect these tools to be useful to future researchers investigating the Apple Watch, we publish them at \url{https://github.com/seemoo-lab/watchwitch-tools}.

\section{Wireless Communication}
\label{sec:communication}

\review{We shortened this section following feedback from reviewer D to put more focus on our other contributions.}

Wireless interfaces form the core of the communication between an Apple Watch and its paired iPhone. A deep understanding of these interfaces is the foundation of our research: Knowing the details of these proprietary and undocumented protocols lets us analyze their security properties, access the Apple Watch with an Android phone, and add new features promoting user privacy and control.

The Apple Watch has two main wireless interfaces, Bluetooth and \mbox{Wi-Fi}. Both interfaces can be used interchangeably to connect to the paired iPhone. The watch and the phone communicate using a common \mbox{Wi-Fi} access point they are both connected to, without using Apple's proprietary \mbox{Wi-Fi} enhancements~\cite{awdl}.

On a high level, all Apple Watch communication is routed over an IPSec tunnel that provides encryption and authentication. This tunnel is established using \gls{ikev2}~\cite{rfcIKEv2} and carries data using \gls{esp}~\cite{rfcESP}.
Both of these protocols are designed to run on top of the \textit{\gls{ip}}. For \mbox{Wi-Fi} connections, this happens transparently. For Bluetooth connections, Apple uses the proprietary \textit{\gls{nrlp}} to carry \gls{ikev2} and \gls{esp} payloads. 

\begin{figure}[!bp]
  \centering
  \includegraphics[width=0.8\linewidth]{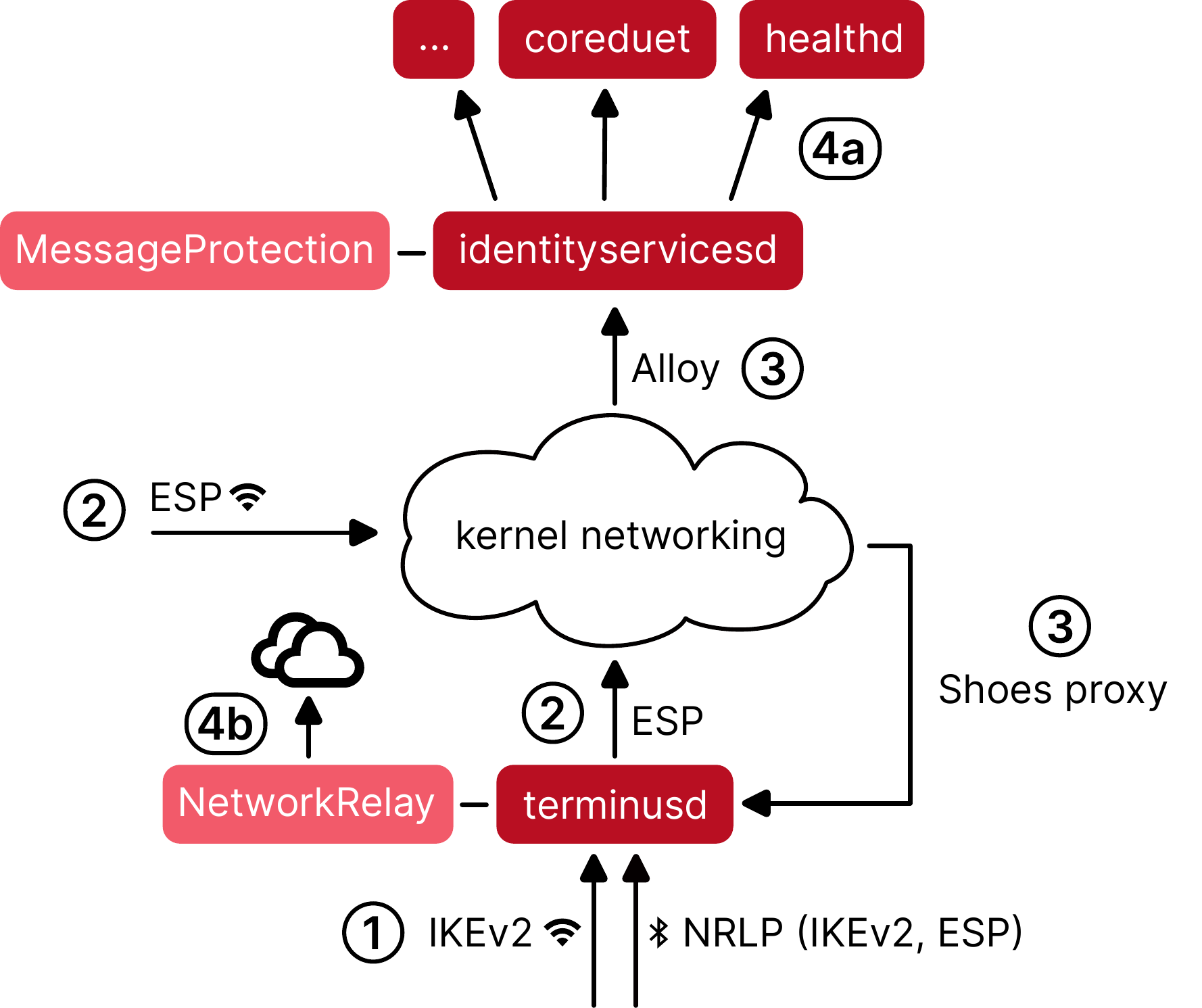}
  \vspace{-0.5em}
  \caption{Watch message handling logic on iOS. Darker cells show daemons involved in the communication, lighter cells show selected frameworks with related functionality.}
  \label{fig:messageflow}
  \Description{Figure fully explained in text.}
\end{figure}

\autoref{fig:messageflow} shows how incoming messages from the Apple Watch are handled on the iPhone. Message handling on the watch should be nearly identical as both devices share a common codebase. However, we focus our investigations on the iPhone side due to the existing jailbreak and tool support for iOS~\cite{dopamine, palerain}.

\begin{enumerate}
    \item \gls{ikev2} packets arrive over the wireless link (Bluetooth or \mbox{Wi-Fi}) and are passed to the \textit{terminus} daemon, which establishes an IPSec tunnel between the devices.
    \item \gls{esp} packets carrying data for the established tunnel arrive and are decrypted by the kernel networking stack. For Bluetooth connections, \gls{esp} packets arrive in \gls{nrlp} payloads, and the terminus daemon forwards them to the kernel.
    \item Decrypted \gls{tcp} packets are passed on to their destinations---the terminus daemon for Internet-bound traffic using the \textit{Shoes} proxy and the identity services daemon for device-to-device traffic using \textit{Alloy}.
    \item[(4a)] For Alloy messages, the identity services daemon performs additional decryption using the \emph{MessageProtection} framework if necessary and passes messages on to their destination services via \gls{xpc} based on the message topic.
    \item[(4b)] For Internet sharing traffic, the terminus daemon passes packets to their intended remote host using the phone's Internet connection.
\end{enumerate}

\begin{figure}[!bp]
  \vspace{-1em} 
  \centering
  \includegraphics[width=\linewidth]{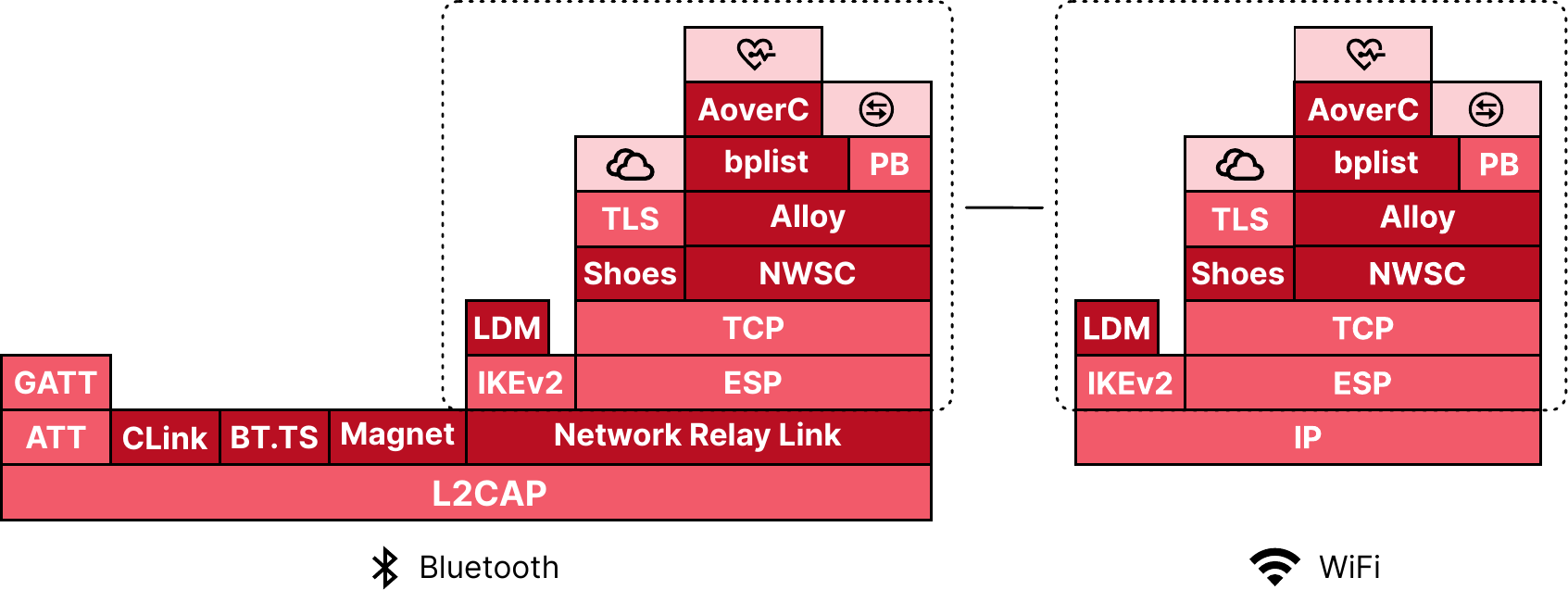}
  \vspace{-0.5em}
  \caption{Protocol stack for Bluetooth and \mbox{Wi-Fi} connections. Cells with a lighter background are standardized open protocols/formats; darker cells are proprietary and largely undocumented. \textit{PB} is \textit{ProtoBuf}. Icons represent health data (heart), Internet access (clouds), and local messages (arrows).}
  \label{fig:protocolstack}
  \Description{The Bluetooth protocol stack consists of L2CAP as a base layer, with GATT/ATT, BT.TS, Magnet, and NRLP building on top off it. The Wi-Fi base layer consists only of IP. The protocols operating on these base layers are identical for both cases: IKEv2 (with LDM) and ESP, on which TCP carries NWSC and Shoes traffic. Shoes in turn carries TLS. NSWC carries Alloy, which carries PB and bplist payloads, which contain local messages or A-over-C payloads containing health data.}
\end{figure}

Regardless of the wireless link technology used, the higher-level protocols remain unchanged, as shown in \autoref{fig:protocolstack}.

\subsection{NRLP \& Magnet}
\label{sec:nrlp}

The foundation of the protocol stack for Bluetooth connections is the \textit{\acrfull{nrlp}}. \gls{nrlp} forms a transport that carries multiple higher-level protocols over \gls{l2cap} channels. It appears to be used exclusively for communication with the Apple Watch and is handled by the terminus daemon. 

\gls{nrlp} utilizes dynamically allocated \gls{l2cap} channels, which are negotiated using Apple's \textit{Magnet} protocol. Magnet was first documented by Heinze et~al.~\cite{toothpickerPaper}. We expand on their work with a detailed list of message types in Appendix~\ref{appendix:magnet}.

We mainly observe \gls{nrlp} carrying \gls{esp} and \gls{ikev2} payloads but, perhaps for historic reasons, other protocol types are supported, including plain \gls{ip} and an echo service. We provide more detail on \gls{nrlp} message types and structures in Appendix~\ref{appendix:nrlp}.

\subsection{IKEv2 \& ESP}
\label{sec:ikeesp}

While Magnet and \gls{nrlp} are used exclusively for Bluetooth-based connections, \gls{ikev2} and \gls{esp}---as well as all the protocols building on top of them---are used for both Bluetooth and \mbox{Wi-Fi} connections with almost no modification. As both protocols are designed to be used with the Internet Protocol, they can be used as-is in the \mbox{Wi-Fi} case. The lower-level \gls{nrlp} provides a Bluetooth compatibility layer for using \gls{ip}-based protocols in this context.

\gls{ikev2}, as used here by Apple, conforms largely to the standard described in RFC~7296~\cite{rfcIKEv2}: It serves as a mutually authenticated key exchange at the start of every connection that establishes ephemeral secrets that will secure all further communication.

Payload data for higher-level protocols will then be encrypted using the negotiated secrets and cryptographic algorithms using \gls{esp}, which Apple uses as described in RFC~4303~\cite{rfcESP}. We list the ciphers used for \gls{ikev2} and \gls{esp} in Appendix~\ref{appendix:crypto}.

\subsubsection{Data Protection Classes}

Apple uses the concept of \textit{data protection classes} to separate data into different sensitivity levels~\cite{dataProtectionClasses}. 

These data protection classes are distinguished by when their decryption keys are available: Class A keys are only accessible when the iPhone is currently unlocked, whereas class C keys are available continuously once the phone has been first unlocked after booting, and class D keys are always available.

The watch opens two separate IPSec tunnels, using the phone's class C and D public keys respectively. The resulting tunnels are used for different kinds of data: The lower-security class D tunnel is used to provide Internet sharing and backup watch settings, which will also work when the paired iPhone has not yet been unlocked after booting. Most other traffic uses the class C tunnel instead, which is only available once the phone has been unlocked.

Health data sent by the watch uses the highest protection class A, for which no dedicated IPSec tunnel exists. Instead, class A data is carried over the class C tunnel with additional \textit{A-over-C} encryption applied. We discuss this encryption in more detail in \autoref{sec:aoverc}.

\subsubsection{Notify Payloads}

In a deviation from `plain' \gls{ikev2}, Apple extensively uses vendor-specific \textit{Notify} payloads. These payloads can be included in any \gls{ikev2} message and carry arbitrary, vendor-specified data typically used for signaling purposes. 
Apple uses these payloads to communicate device names, software versions, and IPv6 addresses used to set up routing for the IPSec tunnel. Beyond that, they also build an entirely separate protocol operating on top of custom \gls{ikev2} notify payloads centered around so-called \textit{\glspl{ldm}}.

\glspl{ldm} have their own protocol header and carry a number of \gls{tlv} encoded substructures. Most interesting of these are the types \textit{Update\mbox{Wi-Fi}Address\-IPv4} and \textit{-IPv6}, which are used by the watch and the phone to discover their \mbox{Wi-Fi} \gls{ip} addresses when connecting over Bluetooth. During the initial connection, or as part of an \gls{ikev2} keepalive when a device's address changes, both parties share their local IP addresses and ports on which they accept \gls{ikev2} connections over \mbox{Wi-Fi}. 

Other \glspl{tlv} can signal that a connection has restarted or negotiate preference for a certain link type (\mbox{Wi-Fi} or Bluetooth). The complete list of supported \glspl{tlv}, alongside other custom notify types and the \gls{ldm} header structure, can be found in Appendix~\ref{appendix:ike}.

\subsection{Alloy}
\label{sec:alloy}

The Alloy protocol is the heart of all communication between the Apple Watch and the phone. Operating on top of the IPSec tunnel, it forms the main messaging bus that delivers messages to many different services. As such, it is rather complex: Alloy defines a total of 54 message types and 231 message topics, which it delivers to and from more than 150 different binaries on the iOS side.

Alloy uses several long-standing \gls{tcp} connections---one control channel and multiple data channels, distinguished by data protection class and urgency. The \textit{identity services daemon} handles all these connections, listening on port 61315 (for the control channel) and 61314 (for most data channels). 

To open multiple distinct data channels using a single server port, Apple uses yet another proprietary protocol called \textit{\gls{nwsc}}. \gls{nwsc} is of little importance for the operation of the other protocols beyond negotiating \gls{tcp} connections, which is why we do not discuss it in detail.

\subsubsection{Control Channel}

Once a control channel has been established, both devices send \textit{Hello} messages, containing version information, device identifiers, and support for various features. Immediately after the Hello message, each device sends a \textit{SetupChannel} message for every data channel it wants to open with the remote device. These messages contain \gls{tcp} port numbers,  \glspl{uuid} used to refer to the channel being established, 
as well as a service identifier composed of \textit{account}, \textit{service}, and \textit{name} parts. All channels we observe share the common account \textit{"idstest"} and service \textit{"localdelivery"}. The name components differ from channel to channel---some values include \texttt{UTunDelivery\b-Default\b-Urgent\b-C} or \texttt{UTunDelivery\b-Default\b-Default\b-D}, where C or D denote data protection classes.

\begin{table}[!tp]
\caption{Types of Alloy control channel messages.}
\small
\begin{tabular}{|r|p{2.8cm}|l|r|p{3.4cm}|}
\cline{1-2} \cline{4-5}
\textbf{\#} & \textbf{Name}         & \textbf{} & \textbf{\#} & \textbf{Name}                \\ \cline{1-2} \cline{4-5} 
\cellcolor{witchred}\textcolor{white}{1}           & Hello                 &  & \cellcolor{witchred}\textcolor{white}{7} & FairplayHostSessionInfo      \\ \cline{1-2} \cline{4-5} 
\cellcolor{witchred}\textcolor{white}{2}           & SetupChannel          &  & \cellcolor{witchred}\textcolor{white}{8} & FairplayDeviceInfo           \\ \cline{1-2} \cline{4-5} 
\cellcolor{witchred}\textcolor{white}{3}           & CloseChannel          &  & \cellcolor{witchred}\textcolor{white}{9} & Fairplay\-Device\-Session\-Info    \\ \cline{1-2} \cline{4-5} 
\cellcolor{witchred}\textcolor{white}{4}           & CompressionRequest    &  & \cellcolor{witchred}\textcolor{white}{10} & OTR\-Negotiation\-Message        \\ \cline{1-2} \cline{4-5} 
\cellcolor{witchred}\textcolor{white}{5}           & CompressionResponse   &  & \cellcolor{witchred}\textcolor{white}{11} & Encrypt\-Control\-Channel        \\ \cline{1-2} \cline{4-5} 
\cellcolor{witchred}\textcolor{white}{6}           & Setup\-Encrypted\-Channel &  & \cellcolor{witchred}\textcolor{white}{12} & SuspendOTR\-Negotiation\-Msg \\ \cline{1-2} \cline{4-5} 
\end{tabular}
\label{tab:alloyCCTypes}
\end{table}

After all desired channels have been established, the control channel has served its main purpose. While channels may be closed and reestablished, there is little control channel activity after the initial connection setup. \autoref{tab:alloyCCTypes} shows a list of supported control channel messages, including many apparently deprecated commands.

\subsubsection{Data Channels}

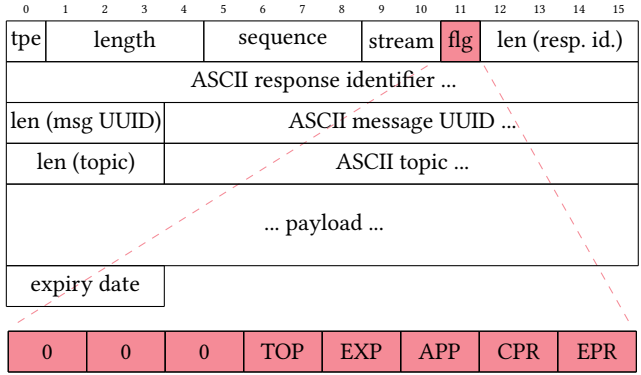
\begin{figure}[!tp]

    \raggedleft
    \begin{tikzpicture}

        \draw[dashed, color=witchlightred, draw opacity=1.0] (1.4, 1.25) -- (-4.1, -2.2);
        \draw[dashed, color=witchlightred, draw opacity=1.0] (2.15, 1.25) -- (4.1, -2.2);

        \node[inner sep=0pt] (fm) at (0,0) {
            \begin{bytefield}[bitwidth=1.66em]{16}
                \bitheader{0-15} \\
                \bitbox{1}{tpe} & \bitbox{4}{length} & \bitbox{4}{sequence} & \bitbox{2}{stream} & \bitbox{1}[bgcolor=witchlightred]{flg} & \bitbox{4}{len (resp. id.)} \\
                \bitbox{16}{ASCII response identifier ...} \\
                \bitbox{4}{len (msg UUID)} & \bitbox{12}{ASCII message UUID ...} \\ 
                \bitbox{4}{len (topic)} & \bitbox{12}{ASCII topic ...} \\ 
                \wordbox{2}{... payload ...} \\
                \bitbox{4}{expiry date}
            \end{bytefield}};
    
        \node[inner sep=0pt] (ll) at (0,-2.5) {
            \begin{bytefield}[bitwidth=3.3em, bgcolor=witchlightred]{8}
                \bitheader{7-0} \\
                \bitbox{1}{0} & \bitbox{1}{0} & \bitbox{1}{0} & \bitbox{1}{TOP} &
                \bitbox{1}{EXP} & \bitbox{1}{APP} & \bitbox{1}{CPR} & \bitbox{1}{EPR}
            \end{bytefield}};
    
    \end{tikzpicture}
    
    \caption{
    Common byte structure of Alloy application data messages. Topic, topic length, and expiry date are only present if the matching flags are set. Flags are TOP: hasTopic, EXP: has\-Expiry\-Date, APP: wants\-App\-Ack, CPR: compressed, EPR: expects\-Peer\-Response. 
    }

    \Description{Alloy application data messages consist of: A type byte, four length bytes, four sequence bytes, a two-byte stream identifier and a flag byte, the lower five bits of which encode flags TOP, EXP, APP, CPR, and EPR. This is followed by length-prefixed response identifier, message UUID and topic strings. After this follows the payload data and a four-byte expiry date.}
    
    \label{bytes:alloy:commonMessage}
\end{figure}

Alloy data channels carry application data between the watch and the iPhone. A message containing this data can be either a generic \textit{DataMessage}, a \textit{DictionaryMessage} containing a dictionary encoded as a binary \textit{plist}, a \textit{ProtobufMessage} containing data encoded using Google's \textit{Protocol Buffers}~\cite{protobuf}, or a \textit{ResourceTransferMessage} containing data fragments of a file transfer.\footnote{These four of about 40 Alloy message types carry the bulk of all communication. Further messages are used for signaling; most appear to be deprecated or unused.} All message types share the format shown in \autoref{bytes:alloy:commonMessage}. Message payloads may optionally be compressed using gzip/deflate~\cite{rfcGzip}, although the \textit{compressed} flag does not actually indicate if compression was applied and appears to be deprecated.

Every data message has a \textit{stream} associated with a \textit{topic}. The topic determines which service on the device will receive and handle the message---health data, for example, is directed to the \textit{health} daemon using the topic \texttt{com\b.apple\b.private\b.alloy\b.health\b.sync\b.classc}. The first message sent for a given stream includes the topic explicitly, while all following messages omit the topic string that can now be inferred from the stream id. Each message also carries its own \textit{message \gls{uuid}}. If a message is a direct response to a prior message, it will set its \textit{response identifier} to the \gls{uuid} of that message. Some messages also contain an \textit{expiry date}. If a message is received after its expiry date, the receiver discards it and acknowledges it with an \textit{ExpiredAck} instead of an \textit{Ack} message.

The receiving service, such as the health daemon, can define its own data format for the payload carried by Alloy. Most services use either binary property lists (bplists) or ProtoBuf encoded data.

\subsection{Additional Data Protection (A-over-C)}
\label{sec:aoverc}

\begin{figure}[!tp]
  \centering
  \includegraphics[width=0.7\linewidth]{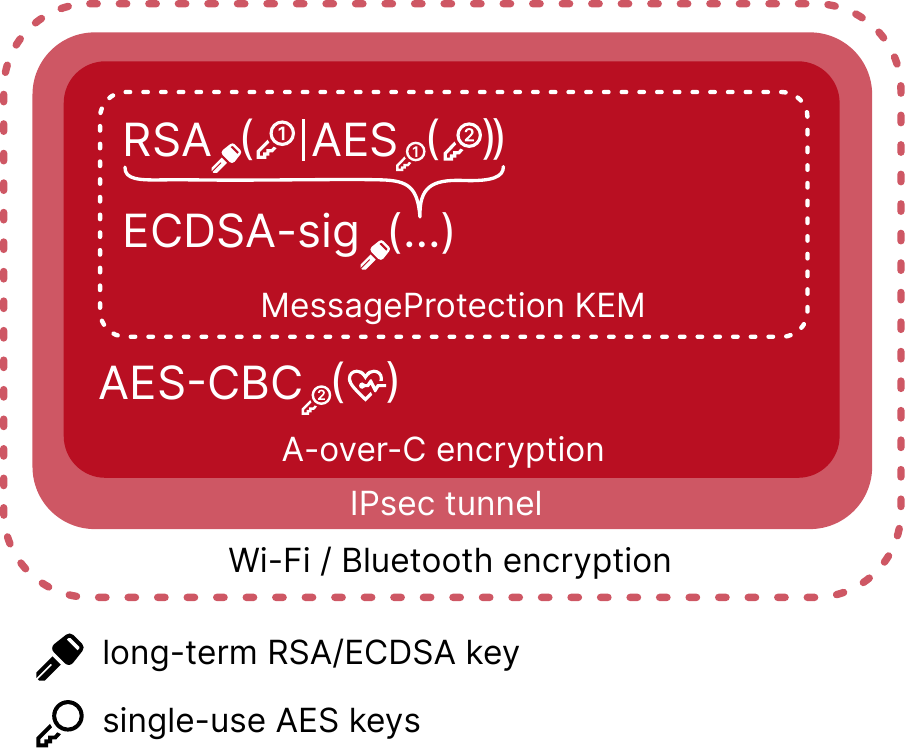}
  \caption{A-over-C layer with surrounding encryption layers.}
  \label{fig:aoverc}
  \Description{There are three nested encryption layers: The outermost is the Wi-Fi / Bluetooth link encryption. Within this is the IPSec tunnel. A-over-C encryption takes place within this tunnel, and consists of a key encapsulation part using the MessageProtection framework, and the actual encryption of payload data using AES-CBC with a single-use key.}
\end{figure}

So far, we have seen how Alloy is used to transport data for protection classes C and D over their respective IPSec tunnels. Meanwhile, health data collected by the Apple Watch falls into the highest protection class A. Since there is no dedicated tunnel for this class, Apple applies additional protection to these messages on a per-message basis using what is referred to as \textit{A-over-C} encryption. 
Plaintext messages are encrypted using a class A key and then delivered via the class C tunnel. On the iPhone the identity services daemon decrypts them and forwards plaintext messages to the destination services as usual. Understanding the details of A-over-C is crucial for our ability to judge its security, which we will discuss in \autoref{sec:security}.
\autoref{fig:aoverc} illustrates the steps performed to encrypt a given message plaintext $p$:\footnote{The encryption steps performed by MessageProtection are simplified below for the case in which $c_1$ is no longer than 114 bytes, which is the case for A-over-C.}

\begin{enumerate}
    \item Choose random 128-bit ephemeral keys $k_1,k_2$
    \item Encrypt $p$ using AES in \gls{cbc} mode with a zero IV and key $k_2$: $sed = AES-CBC_{k_2}(p)$
    \item Encrypt $k_2$ using AES in counter mode with key $k_1$: $c_1 = AES-CTR_{k_1}(k_2)$
    \item Encrypt $k_1$ and $c_1$ using RSA-OAEP with the receiver's long-term public key $pk_r$: $c_2 = RSA-OAEP_{pk_r}(k_1 || c_1)$
    \item Sign the resulting ciphertext using the sender's ECDSA long-term private key $sk_s$: $s = ECDSASign_{sk_s}(c_2)$
    \item Encode $s$ and $c_2$ as $ekd = version || len(c_2) || c_2 || len(s) || s$
    \item Return a \textit{DataMessage} with a bplist-encoded dictionary containing $ekd$ and $sed$.
\end{enumerate}

The encapsulated key $ekd$ is computed using the \textit{MessageProtection} framework, which uses the same cryptography as Apple's iMessage---which has previously been analyzed by Garman et~al.~\cite{imessageAttack}.

\subsection{Health Data Synchronization}
\label{sec:healthsync}
In this section we discuss the synchronization of health data as a service that powers a core feature of the Apple Watch and carries private and sensitive data. Many other services share the same vocabulary of techniques and encodings used.

When the watch has new health samples available, such as new heart rate measurements, it sends them in an Alloy \textit{DataMessage} with the topic \texttt{com\b.apple\b.private\b.alloy\b.health\b.sync\b.classc}. The message receives additional A-over-C encryption as health data falls into the most sensitive data protection class A. On reception, the phone decrypts the A-over-C ciphertext once it is unlocked and has access to the corresponding class-A keys. As for all Alloy messages, the message is also protected by the IPSec tunnel in transit---A-over-C merely forms an additional encryption layer.

\begin{figure}[!bp]
  \centering
  \includegraphics[width=\linewidth]{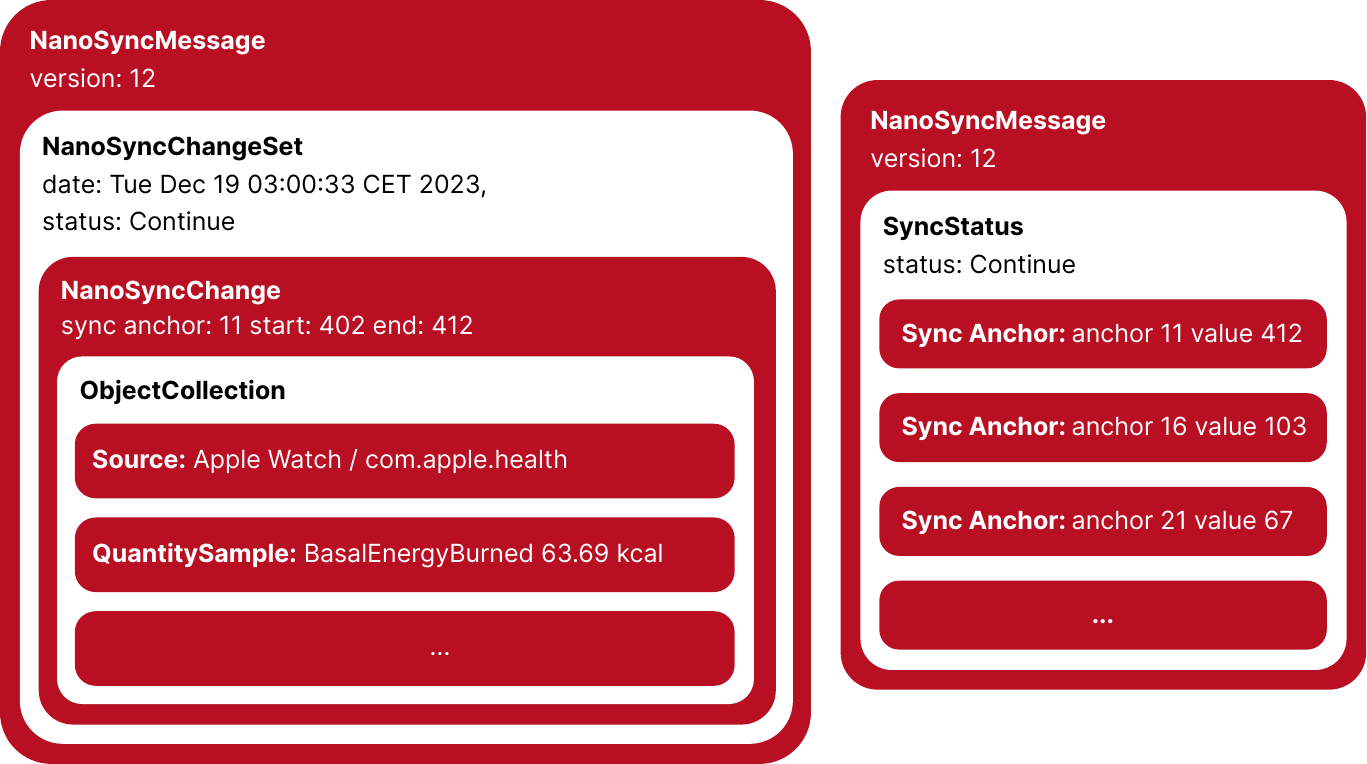}
  \caption{A simplified illustration of a NanoSync message containing new health samples from the watch (left) and a reply message acknowledging receipt of these changes (right).}
  \label{fig:nanosync}
  \Description{On the left: A NanoSyncMessage containing a NanoSyncChangeSet with status "Continue". Within the change set is a NanoSyncChange containing objects of type "Quantity Sample". The change has associated start and end sync anchors. Contained in the change is an ObjectCollection. The object collection has a source and provenenace attribute and in turn contains the quantity samples. On the right: A NanoSyncMessage containing a SyncStatus carrying a set of SyncAnchors acknowledging received changes.}
\end{figure}

The resulting plaintext is forwarded to the health daemon, which decodes the ProtoBuf payload into a \textit{NanoSync} message (\autoref{fig:nanosync}). NanoSync is a lightweight abstraction layer that synchronizes \textit{SQLite} databases on the watch and the phone. The protocol is centered on the notion of \textit{changes}: Each change contains a collection of \textit{samples} of the same type to be inserted into the database. A sample may be a single heart rate measurement or the energy burned by the user over five minutes. Removal of samples is handled similarly: A \textit{change} with a collection of \textit{deleted samples} instructs the health daemon to mark the referenced samples as deleted. 

Changes are ordered by their \textit{sync anchors}. A sync anchor is a counter used for a particular domain of the database that is incremented with every change. This method allows the receiving device to apply updates in the correct order, determine if changes are missing, and acknowledge receipt of new data. After receiving a NanoSync message, it will respond with an updated list of its local sync anchors until the sending device communicates that it has no further changes and the synchronization is complete. We show two typical messages in \autoref{fig:nanosync}.

\subsection{Shoes}
\label{sec:shoes}

The Apple Watch can share a paired phone's cellular Internet access, thus allowing Internet-enabled features on the watch while on the go.
Internet sharing differs from other services in that it does not use Alloy as a messaging bus. 
The watch uses a protocol referred to as \textit{Shoes} to open connections to the Internet via the iPhone. The phone then internally uses components of a SOCKS proxy server to forward traffic from the watch to the destination host and back~\cite{socks}.

\begin{figure}[!tp]
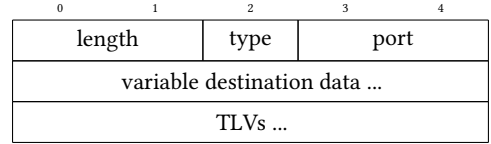

    \centering
    \begin{bytefield}[bitwidth=4em]{5}
        \bitheader{0-4} \\
        \bitbox{2}{length} & \bitbox{1}{type} & \bitbox{2}{port} \\
        \wordbox{1}{variable destination data ...} \\
        \wordbox{1}{ TLVs ... }
    \end{bytefield}
    \caption{A generic Shoes request. Destination data is dependent on the request type (hostname, IPv4, IPv6, bonjour).}
    \Description{A Shoes request starts with a two-byte length field and a single type byte. This is followed by a two-byte destination port and variable length destination host data. The request is concluded by variable length \gls{tlv}-encoded values.}
    \label{bytes:shoes:request}
\end{figure}

\begin{figure}[!tp]
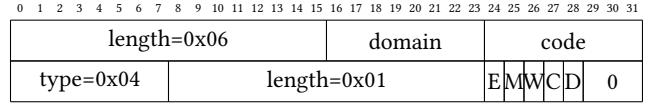

    \centering
    \begin{bytefield}[bitwidth=0.83em]{32}
        \bitheader{0-31} \\
        \bitbox{16}{length=0x06} & \bitbox{8}{domain} & \bitbox{8}{code} \\
        \bitbox{8}{type=0x04} & \bitbox{16}{length=0x01} & \bitbox{1}{E} & \bitbox{1}{M} & \bitbox{1}{W} & \bitbox{1}{C} & \bitbox{1}{D} & \bitbox{3}{0}
    \end{bytefield}
    \Description{A Shoes reply starts with a two-byte length field. The length of the reply is always six. It is followed by single-byte "domain" and "code" fields. After those follows a \gls{tlv} section containing a single \gls{tlv} of type four (encoded in a single type byte) and length one (encoded in a two-byte length field). The value byte of the \gls{tlv} contains the flag bits E, M, W, C, and D as described in the caption, followed by three zero bits.}
    \caption{Bit structure of a Shoes reply. The second word is a \textit{network info} \gls{tlv} field. Flag names: (E) expensive, (M) cellular connection, (W) \mbox{Wi-Fi} connection, (C) constrained connection, (D) connection denied.}
    \label{bytes:shoes:reply}
\end{figure}

To open a new connection, the watch sends a \textit{Shoes request} (\autoref{bytes:shoes:request}) to the phone on \gls{tcp} port 62742. This request includes the desired destination host and port, usually in the form of a hostname or \gls{ip} address. Typically, the watch also includes optional \glspl{tlv} containing the name of the requesting process and a set of flags indicating in which network conditions the request should be fulfilled. The watch can, for example, specify that a large file transfer should only be completed when connected to \mbox{Wi-Fi}, while an expensive cellular connection may be used to fetch weather data.

On the phone, the terminus daemon receives these requests and checks the specified conditions against the phone's current network connection. It then responds with a \textit{shoes reply} (\autoref{bytes:shoes:reply}) communicating whether the request was accepted and what the phone's current connection status is. On success, the phone will forward any following bytes received on the connection to the destination host, sending replies to the watch the same way. From this point onwards, the watch can treat the connection as a regular \gls{tcp} connection to the destination host. It typically continues to open a \gls{tls} session with the remote server, but would be free to continue with other protocols as well.

\section{W\MakeLowercase{atch}W\MakeLowercase{itch} for Android}
\label{sec:reimplementation}

With the understanding of the protocol stack we gained from reverse-engineering, we can reimplement these protocols in the \textit{WatchWitch} app to demonstrate that interoperability with Android is practically possible. As part of our reimplementation, we will also extend WatchWitch with several features designed to give users autonomy over their data and better privacy controls. 

\subsection{Design Goals}

One explicit goal of our work is to show that meaningful interoperability between the Apple Watch and third-party smartphones is possible, despite Apple's claims to the contrary. To do so, we reimplement a usable selection of \textbf{essential smartwatch features}, including support for push notifications, Internet sharing, and access to health data. While full support of most or all watch features is far beyond the scope of this work, this already provides value to potential users and proves that there are no fundamental barriers to true interoperability. Our architecture also provides all the required infrastructure to allow technically adept users to implement support for any other services they require.

A common argument against interoperability are security concerns.
The recent US antitrust complaint picks up on this, stating that \textit{"Apple deploys privacy and security justifications as an elastic shield that can stretch or contract to serve Apple’s financial and business interests"}~\cite[p.~12]{antitrustUsa}. To show that interoperability does not necessarily sacrifice security, we therefore \textbf{maintain security} on a level comparable to the security provided by Apple out of the box. We do not circumvent or downgrade any protection mechanisms and complete all encryption and authentication steps as required by the respective protocols using secure key material. To ensure that long-term keys are stored with a level of protection comparable to the system keychain~\cite{keychainPlatformSecurity} used on iOS, we keep our cryptographic secrets in Android's hardware-backed KeyStore~\cite{androidKeystore, androidKeystoreAnalysis}. 

Beyond providing an experience on par with what Apple offers for iOS users, we strive to \textbf{enhance privacy} in our implementation, allowing users to use their smartwatch entirely offline (without connecting to any commercial cloud servers) and putting them in explicit control of their data. We hope to demonstrate that in this way, open interfaces and interoperable devices do not just benefit people with different smartphones but all users across the spectrum in a sort of curb-cut effect. We want our implementation to \textbf{enable users} to use their devices to the fullest, going beyond the at times limited uses intended by the manufacturer.

\subsection{Architecture}

For WatchWitch, we focus on the \mbox{Wi-Fi} part of the Apple Watch communication infrastructure---since the underlying wireless link is transparent to application-level messages, this does not limit the available features. Using standard \mbox{Wi-Fi} makes it easier for us to interact with the watch from an Android app.\footnote{Real Bluetooth connectivity should also be possible using raw \gls{l2cap} sockets on Android. Work on extending WatchWitch to this use case is ongoing.}

\begin{figure}[!tp]
  \centering
  \includegraphics[width=\linewidth]{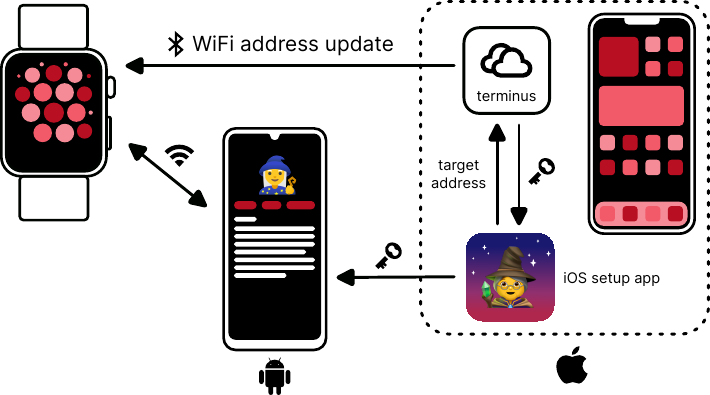}
  \caption{The WatchWitch app in context, showing the Apple Watch and the paired iPhone as well as the Android phone running the app.}
  \label{fig:setup}
  \Description{On the iPhone, our iOS setup app modifies the terminus daemon to inject the target IP address extracts keys from it. It then sends these keys to the Android phone. The terminus daemon sends a Wi-Fi address update notification to the watch via Bluetooth, containing our target address. The watch and the Android phone then communicate over Wi-Fi.}
\end{figure}

Since the watch requires Bluetooth connectivity for the initial setup and \mbox{Wi-Fi} discovery, our setup (shown in \autoref{fig:setup}) includes a jailbroken iPhone that performs the watch setup and hands over communication to the Android phone by sending a \mbox{Wi-Fi} address update. After the setup is completed, we extract the cryptographic long-term keys from the phone and securely transfer them to the Android device using a custom app and tweak.\footnote{We use Theos~\cite{theos} and Cephei~\cite{cephei} to build a tweak and accompanying iOS app that modify the behavior of the terminus daemon and give us access to long-term keys.} At this point, the iPhone is only required to perform the \mbox{Wi-Fi} discovery mechanism occasionally, e.g. after the watch is restarted. It is no longer part of the communication between the watch and the Android phone.

\begin{figure}[!tp]
  \centering
  \includegraphics[width=\linewidth]{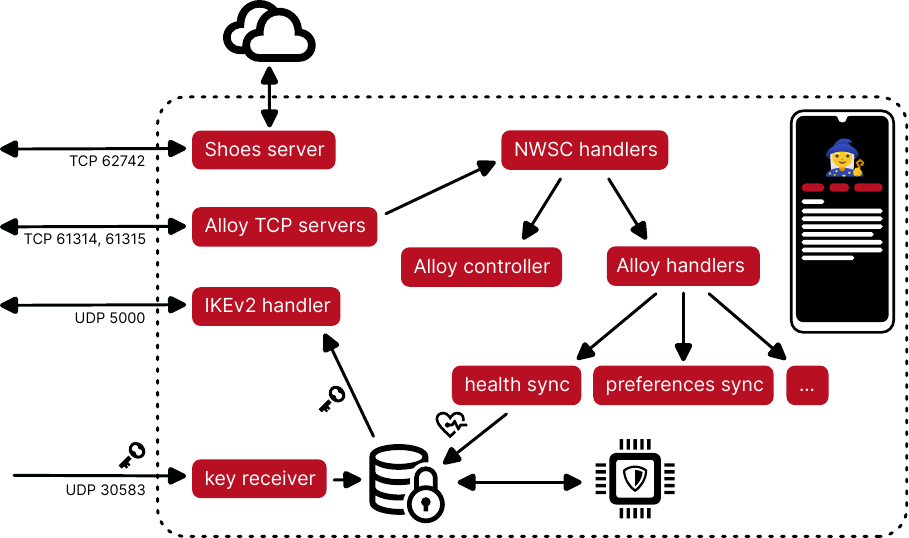}
  \caption{Overview of the internal components of the WatchWitch app, showing how incoming traffic is handled.}
  \label{fig:architecture}
  \Description{The WatchWitch Android app receives traffic in several ways: Extracted keys from the iOS setup app arrive in the key receiver and are stored using Android's TEE hardware protection. IKEv2 traffic from the watch arrives on UDP port 5000. The IKEv2 handler receives keys from storage to handle this traffic. Shoes traffic arrives on TCP port 62742 and is handled by the Shoes server before being forwarded to the Internet. Alloy traffic arrives on TCP ports 61314 and 61315 and is handled in several stages before being passed to service implementations, including health sync and preferences sync. The health sync service stores health data in an encrypted, TEE-backed database.}
\end{figure}

On the Android phone, we replicate the protocol handling we analyzed on the iPhone within our WatchWitch app (see \autoref{fig:architecture}): The phone handles incoming \gls{ikev2} messages to set up an IPSec tunnel, receives packets on that tunnel and forwards them to the Shoes or Alloy handling components. Internet-bound Shoes traffic is forwarded using the phone's network connection, while Alloy messages are handled locally and forwarded to service reimplementations registered for particular message topics. 

Our Android app currently requires root access to the phone to set up the IPSec tunnel and associated routing. We discuss these limitations in more detail in \autoref{sec:limitations}.

\subsection{Features}

By providing implementations of the foundational protocols discussed in \autoref{sec:communication}, we create a base layer to build support for various features. For the initial version of WatchWitch, we focus on three core smartwatch features: Receiving notifications, sharing Internet connectivity, and synchronizing health data.

\subsubsection{Notification Forwarding \& Message Replies for Android Apps}

The push notifications displayed on the Apple Watch---most notably for instant messaging using iMessage and other apps---internally use a service referred to as \textit{bulletin distributor}. The bulletin distributor on the iPhone sends ProtoBuf-encoded messages to the watch, keeping it informed about the state of the notifications present on the iPhone. These messages include information about the source and content of the notification, as well as the actions that can be taken in that context: Acknowledging, snoozing, or dismissing a message, as well as composing a reply on the watch.  

\begin{figure}[!tp]
  \centering
  \includegraphics[width=\linewidth]{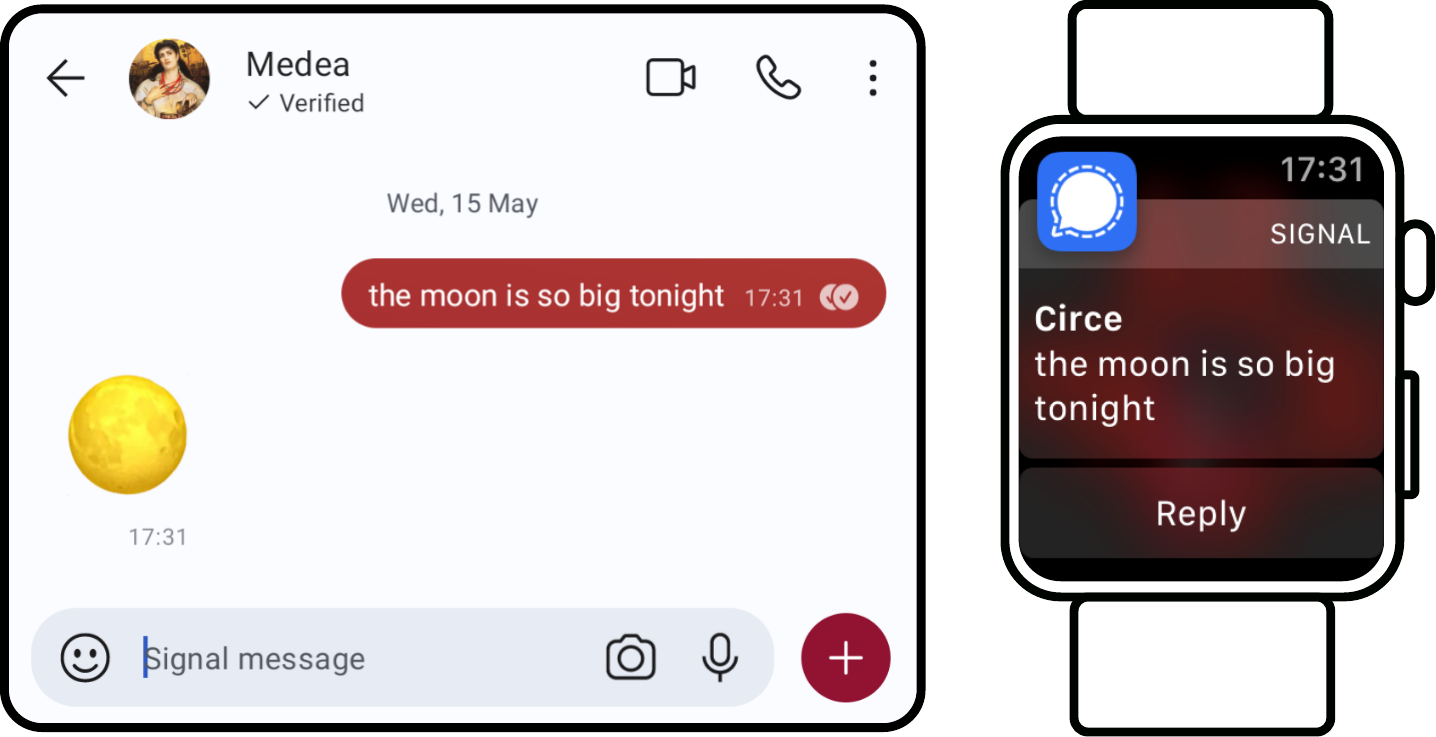}
  \caption{A Signal message sent to the Android phone is displayed on the watch, and the emoji reply sent from the watch received at the sender.}
  \Description{On the left: a screenshot of a Signal chat with a person called Medea, showing the message "the moon is so big tonight" sent to them and them replying with a moon emoji. On the right: a screenshot of the Apple Watch showing an incoming Signal notification from Circe reading "the moon is so big tonight" with a reply button underneath.}
  \label{fig:signal}
\end{figure}

The WatchWitch app can optionally register as a notification listener on Android. If the app then observes an incoming instant messaging notification (e.g. a Signal message), it generates a \textit{bulletin request}, which is sent to the connected watch, instructing it to notify the user. While Apple famously does not allow third-party smartwatches to reply to incoming notifications~\cite[p.~40]{antitrustUsa}, we can use our knowledge of the bulletin distributor combined with Android's Notification Listener Service~\cite{androidNotificationListener} to send messages with reply actions to the watch from our Android phone. Registering a notification listener requires explicit user consent but is possible for regular apps without root privileges. Responding to messages does not require any modifications to the app sending the notification but does rely on it supporting quick reply actions within its notifications. As a demonstration, we include support for receiving and replying to Signal messages in our WatchWitch app, as shown in \autoref{fig:signal}.


\begin{figure}[!tp]
  \centering
  \includegraphics[width=\linewidth]{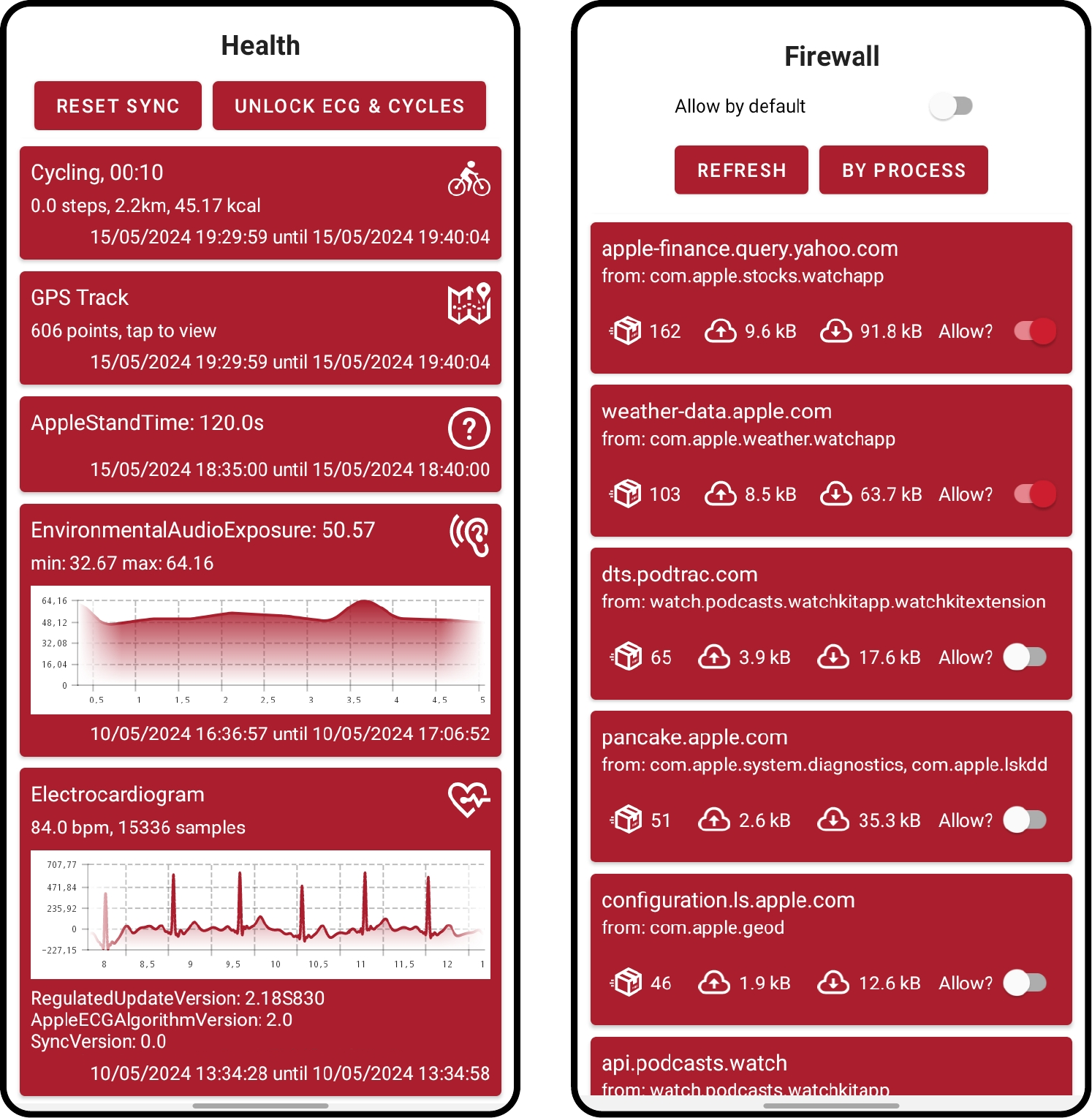}
  \caption{WatchWitch's health log and firewall views.} 
  \Description{On the left: The health view of the WatchWitch app, showing a chronological log of received samples, including an electrocardiogram, a plot of ambient noise, a cycling workout and an associated GPS track. On the right: The firewall view, showing a list of hosts contacted by different processes. Below each host is a summary of how much data was sent to and received from it, as well as a switch that blocks connection to it.}
  \label{fig:screenshots}
\end{figure}

\subsubsection{Internet Sharing \& User-Controlled Firewall}

WatchWitch includes an implementation of \textit{Shoes} (\autoref{sec:shoes}) that responds to connection requests from the watch and forwards traffic using the phone's Internet connection. As the watch prefers to use the phone's connection even when it is connected to a \mbox{Wi-Fi} network, this effectively puts the app into a proxy position for the entirety of the watch's Internet-bound traffic.

With WatchWitch, we can use this position to show users which processes on the watch connect to which hosts, and how much data they transfer---be that for app data, logging, or tracking. Going one step further, we implement a user-controlled firewall that allows users to selectively block connections to certain hosts---or cut off apps from the Internet entirely. This makes it possible to allow connections to the weather \gls{api} required by the watch's built-in weather app, but block all traffic that could be detrimental to user privacy. With the amount of tracking present especially in third-party apps~\cite{third-party-tracking-in-apps}, this presents a powerful tool for users to make fine-grained trade-offs between privacy and functionality.

To see how our firewall is beneficial in a real-world setting, we test it with four watch apps that are, as of May 2024, prominently featured in Apple's watchOS app store: \textit{SmartGym}~\cite{smartgym}, \textit{Pedometer++}~\cite{pedometer}, \textit{Carrot Weather}~\cite{carrotweather}, and \textit{Outcast}~\cite{outcast}. Both SmartGym and Pedometer++ attempt to connect to a remote server as we use them, but continue to function if we completely deny them Internet access. Carrot Weather needs a connection to its Weather \gls{api} to work, but also connects to a second host for what we assume are tracking or analytics purposes. Using host-based blocking, we maintain the the app's functionality while blocking this secondary non-essential connection. We can use a similar approach with the podcast app Outcast: Besides some Apple services, the app connects to its own \gls{api} server, which is required to add new podcast feeds. Once a podcast is added, however, we can block all of these connections, only allowing the app to reach the server hosting the actual podcast. This shows we can improve user privacy in the real world by \emph{(a)} keeping some apps entirely offline and \emph{(b)} only allowing essential hosts for other apps.

\subsubsection{Health Data Synchronization}

With what we learned about the ProtoBuf-based \textit{NanoSync} protocol in \autoref{sec:healthsync}, supporting health data synchronization in WatchWitch is relatively straightforward: We copy the database structure found on the iPhone in a dedicated SQLite database on the Android phone and insert new samples received from the watch using the same \gls{sql} statements used in the health daemon on iOS. 

On the iPhone, health database files are protected using the keychain's data protection, ensuring that they remain encrypted and inaccessible when the device is locked. To achieve a comparable level of security, we take inspiration from how Signal protects its message databases~\cite{signaldatabasesecret}: Using SQLCipher~\cite{sqlcipher}, we encrypt the database on the fly using key material protected by the Android KeyStore. Keys only remain in-memory while the app is running, giving us similar hardware-backed protection as on the iPhone.

WatchWitch can run arbitrary queries against the health database to surface sensor measurements, workouts, etc. to the user. Because we are not bound by the same legal constraints as Apple, we can expand the watch's functionality here: As discussed in \autoref{sec:background:features}, some features are only available in certain geographic regions. Apple enforces this by requiring users to manually enable these features while their iPhone is connected to the cellular network of an allowed country. With WatchWitch, we can generate the required feature unlock messages ourselves and send them to the watch, regardless of our physical location. We demonstrate this for the \gls{ecg} app, but the same approach should also translate to other features. It seems that researchers at Masimo---who are in a patent dispute with Apple over the watch's pulse oximeter---have already used a similar technique to show that watches sold in the US with a disabled oximeter can be reactivated~\cite{oximeterJailbreak}.

Thus, WatchWitch allows us to use the watch's full capabilities as a health and fitness tracker with complete privacy---keeping an encrypted, near-perfect drop-in copy of the iPhone's health database on the phone without any data ever leaving the device. Even if Apple were to send health data directly to the cloud from the watch (which we have no evidence of in current versions), WatchWitch could still block these cloud services using the firewall feature, keeping health data completely on-device. With full database access, we can also give users access to data not usually available in Apple's \gls{ui}, including long-term ambient noise levels and historic GPS tracks (see \autoref{fig:screenshots}). Detailed GPS tracks in particular are usually only stored for a set amount of time on the iPhone~\cite{forensicsHealthdb}.

\subsubsection{Further Features}

The WatchWitch app supports receiving screenshots taken on the watch; it allows users to see which apps are currently running; which alarms are set; and if the watch is muted. Much of this state information is not easily accessible to the user on iPhones: We tap the watch's configuration backup mechanism for access, thus once again showing that we can meaningfully extend the features of the Apple Watch beyond what is intended by Apple and expand user's visibility into their devices.

To developers, WatchWitch provides a simple interface that allows them to add support for more services based on Alloy. This interface abstracts from the lower-level transport protocols, formats, and connection management---from a developer perspective, the new service can simply send and receive arbitrary Alloy messages.

Finally, our app stores full communication transcripts and file transfers received from the watch for later analysis. We provide an Alloy parser that can process these transcripts and may be used to debug or reverse-engineer Alloy communications with ease. 

\subsection{Limitations}
\label{sec:limitations}

\review{As reviewers B and C correctly pointed out, our prototype has limitations. We now more clearly address these limitations, giving reasons for them, and pointing out directions for future research towards true everyday usability.}

In its current form, the setup required to use WatchWitch is rather complicated: Beyond a rooted Android phone and an Apple Watch, it also requires the presence of a jailbroken iPhone on the same \mbox{Wi-Fi} network to bootstrap the connection and does not work with a Bluetooth link. This setup is impractical for day-to-day use.

Allowing users to setup their watch without ever connecting to an iPhone would require a reimplementation of the Bluetooth-based pairing process, of which we do not yet have a sufficient understanding. As part of the pairing, both devices must obtain cryptographic certificates from Apple's servers, which Apple deliberately restricts to their own devices. Bypassing these restrictions, however, is not impossible~\cite{pypush}.

Many of the complications in our setup stem from our position as unprivileged third-party developers. A party with deeper access on either side of the connection (i.e. Apple as manufacturer of the Apple Watch or Google as developer of Android) could make the interoperability process significantly simpler with only small changes to their products.
In its current state, WatchWitch requires root access to instruct the kernel to decrypt incoming \gls{esp} packets. If Google were to extend its existing IPSec \gls{api}~\cite{androidIpsec} to support \gls{esp} in tunnel mode, this would no longer require root privileges. Alternatively, if Apple allowed \gls{udp} encapsulation of \gls{esp} traffic on the watch, we could receive this traffic on Android without special privileges and perform decryption within the app.
We could also circumvent the need for a rooted phone by reimplementing the initial pairing process, using the Bluetooth link for data transfer, and handling the entire IPSec setup ourselves within the app. The pairing process, however, is not well understood. 

As a result of these barriers, the WatchWitch app that we showcase here is merely a proof of concept: Some features require `stealing' a protocol session from the previously connected iPhone, which does not work with the reliability expected from a consumer application. With many different services that are not yet supported, the app may sometimes behave unexpectedly and fail to parse certain messages. The previously mentioned setup requiring multiple phones currently limits the usability of our app. We therefore opt to focus on WatchWitch's technical capabilities rather than providing a fully user-friendly drop-in replacement for Apple's native implementations---especially as such recreations of Apple's \gls{ui} would raise copyright issues. We note, however, that the infrastructure provided by WatchWitch reduces the task of providing a usable replacement for the Health app, e.g., to a simple exercise in \gls{ui} design, as would be the case for any regular Android app.
Given all these challenges, the app runs remarkably well for long periods once connected to the watch. 
Our testing allowed performing workouts and mobile connections with a hotspot for over two hours. At home, we maintained an active connection for over 24 hours. 

We are actively working to address these limitations as best as we can given the restrictions placed on us by Apple and Google as device manufacturers: In the future, we would like to extend WatchWitch to work on the Bluetooth layer, eliminating the need for an iPhone in the loop as well as a shared \mbox{Wi-Fi} network. Advances in this direction would also allow us to focus on real usability, as well as provide interoperability in the opposite direction: allowing the use of third-party smartwatches on iOS with the same level of deep integration as the Apple Watch.

\section{Security Analysis}
\label{sec:security}

\review{We significantly expanded this section to include a more detailed attacker model, including attacker capabilities and motivations as well as a comparison to the competing fitbit ecosystem.}

While reverse engineering the Apple Watch protocol stack, we have familiarized ourselves with its security architecture.

Overall, Apple's multi-layer encryption approach offers strong protection for all application traffic---the IPSec encryption layer alone should prevent most real-world attacks on the watch. This distinguishes the Apple Watch from other smartwatches that rely exclusively on link-layer encryption.

The parts of the protocol stack where Apple veers from established standards, however, are haunted by legacy support and questionable decisions: We find violations of common cryptographic practices, malleable encryption, and unintended interactions between standard and non-standard protocols. Beyond these issues, the protocol stack contains large amounts of complexity, much of it due to legacy versions and deprecated features. This presents a significant and previously unexplored attack surface.

\subsection{Threat Model}

\review{We added additional clarifications for both attacker models, explaining how they correspond to plausible real-world threats.}

We analyze the security of the Apple Watch in an everyday scenario, during which the watch is worn throughout the day in public. The watch connects to the phone to exchange messages over Bluetooth or shared \mbox{Wi-Fi} networks. We do not consider the setup of a new watch as part of this scenario, as this step is performed only once and likely happens in a private space, making it difficult for attackers to get into radio range at exactly the right time. The initial Bluetooth pairing of the Apple Watch has been briefly analyzed in prior research and follows best practices~\cite{watchPairingSecurity}.

Our focus lies on Apple's proprietary protocols, as these parts distinguish the Apple Watch from other devices.
In this section we consider Dolev-Yao attackers~\cite{dolevYao} with the ability to read, modify, and inject messages at will. We assume that all attackers are within radio range of both devices and are members of the same wireless network. With these assumptions, attackers attempt to interfere with the Apple Watch with one of the following goals:

\setlength{\leftskip}{1.75em}  
\hspace*{-1.7em}\textbf{Confidentiality}: Gain access to private data, especially sensitive health information.

\hspace*{-1.7em}\textbf{Integrity}: Place or manipulate data on the watch or phone.

\hspace*{-1.7em}\textbf{Availability}: Disrupt communication between watch and phone or delete data on either device.

\setlength{\leftskip}{0em}  

A peculiarity of the Apple Watch's protocol stack is its highly layered nature: The \mbox{Wi-Fi} or Bluetooth transport layer \textit{may} already provide some security, the IPSec tunnels provide a first and tested layer of encryption, and the A-over-C layer provides another additional encryption layer. With a simple network attacker, we could not analyze these nested layers independently. We therefore simulate attackers that have already broken some of the protection layers by giving them access to the corresponding keys. While this key-access is artificial, these attackers still model plausible real-world attacks: For the \mbox{Wi-Fi} / Bluetooth layer (Attacker~1), this matches the capabilities of someone with access to a Bluetooth or \mbox{Wi-Fi} vulnerability, or someone with access to the same shared \mbox{Wi-Fi} network. Attackers able to bypass the IPSec layer (Attacker~2) might be able to do so after gaining physical access to a locked phone and extracting parts of its memory~\cite{cellebriteLeak}.

\paragraph{Attacker~1} This attacker models a broken transport layer, which is plausible since there have been many generic attacks against \mbox{Wi-Fi} and Bluetooth in the past~\cite{wifiDragonblood, wifiKrack, bluetoothCryptoAnalysis}, as well as Apple-specific security issues~\cite{frankenstein,toothpickerPaper,wangyu_wifi}. Furthermore, transport layer encryption does not provide any protection against attackers if the devices communicate over a \mbox{Wi-Fi} network that the attacker can join, such as open \mbox{Wi-Fis} or \mbox{Wi-Fis} with a static WPA2 passphrase.  
The main attack surface for Attacker~1 is the IPSec layer: Extracting or manipulating data requires circumventing the encryption and authentication provided by the \gls{vpn} tunnel. We assume IPSec in its well-known standard form is secure as it has been formally analyzed~\cite{ipsecAnalysisThesis} and the chosen cryptographic primitives are secure (see Appendix~\ref{appendix:crypto}). 
We focus on the security implications of Apple's deviations from the standard in \autoref{sec:security:ike}.

\paragraph{Attacker~2} This attacker has access to IPSec key material, giving them `legitimate' access to most of the watch's communication. Such a severe break in the IPSec security layer is unlikely---but this attacker corresponds to a scenario where adversaries briefly gain physical access to the iPhone while it is locked---such as in a border control setting. In this state, class C and D key material is available and could be extracted. Companies like Cellebrite or Magnet Forensics offer such services commercially~\cite{cellebriteLeak, magnetGraykey}. Key material for class A, however, should remain secure, and attackers should not be able to decrypt or modify traffic protected with these keys.
Attacker~2, having access to the IPSec tunnel, already has a large amount of control over the system. For example, it can disrupt communication by continuously resetting the watch. The only remaining trust boundary for this attacker is gaining access to class-A protected data. The main attack surface is the A-over-C protocol, which we analyze in \autoref{sec:security:forging_health_data}.

\subsection{Insecure IKEv2 Extensions}
\label{sec:security:ike}

Where Apple extends the proven \gls{ikev2} standard with their own \gls{ldm} protocol, they fail to account for the fact that such payloads may be included in unencrypted, unauthenticated \gls{ikev2} messages. As the first messages of any \gls{ikev2} handshake are---by necessity---unencrypted, the \gls{ikev2} standard allows \textit{notify} payloads in unencrypted contexts. While Apple only uses custom notify payloads once encryption is established, they never check if the payload was received in an encrypted message. They also continue to accept unencrypted messages after encryption is successfully established.

\begin{figure}[!tp]
  \centering
  \includegraphics[width=\linewidth]{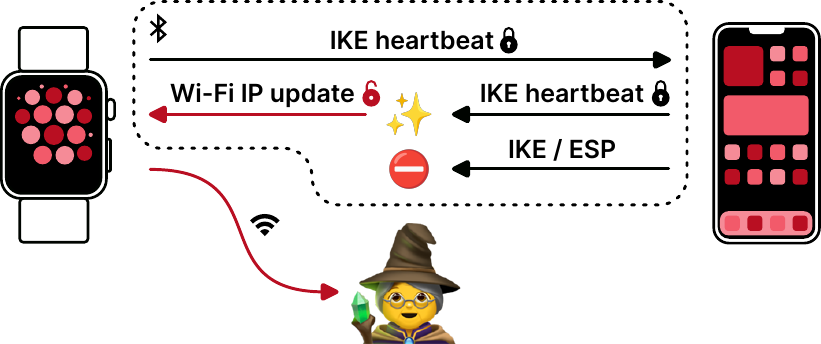}
  \caption{Attacking the Apple Watch by replacing an \gls{ikev2} heartbeat with a forged, unencrypted \mbox{Wi-Fi} IP address update and redirecting \mbox{Wi-Fi} traffic to the attacker.}
  \label{fig:wifiUpdateInject}
  \Description{The attacker replaces an encrypted IKEv2 heartbeat sent in an established Bluetooth connection with an unencrypted IKEv2 message containing a Wi-Fi IP update and blocks any further Bluetooth communication. The watch then connects to the attacker using Wi-Fi in an attempt to reconnect.}
\end{figure}

Thus, an attacker with the ability to inject Bluetooth or \mbox{Wi-Fi} packets (Attacker~1) can send forged Link Director Messages---for example to manipulate or jam the \mbox{Wi-Fi} discovery mechanism discussed in \autoref{sec:ikeesp}, as shown in \autoref{fig:wifiUpdateInject}.
While this vulnerability allows attackers to redirect watch traffic to an attacker-controlled device, it does not give them the ability to interact meaningfully with the watch without the required cryptographic keys. Nonetheless, this oversight by Apple opens up all of the complexity of their custom notify payloads to unauthenticated attackers---a similar attack could, for example, redirect Shoes proxy traffic to unintended or malicious destinations. This could be avoided by only accepting custom notify payloads in encrypted and authenticated contexts.

\subsection{Forging Health Values in A-over-C}
\label{sec:security:forging_health_data}

 The A-over-C protocol (\autoref{sec:aoverc}) uses the same cryptography as iMessage, which has been shown to be vulnerable in the past and has historically employed short key lengths and obscure protocol design~\cite{imessageAttack}. Beyond that, A-over-C awkwardly composes iMessage-based encryption with other primitives, which leaves the payload data entirely unauthenticated---data carried by A-over-C is encrypted using only unauthenticated AES-\gls{cbc} encryption (see \autoref{sec:aoverc}). A drawback of the \gls{cbc} mode of operation is its malleability: If an attacker flips bits in a ciphertext block $c_1$ (creating $c_1' = c_1 \oplus x$), this will cause the corresponding plaintext block $p_1'$ to be corrupted and essentially random. However, the \textit{following} ciphertext block $c_2$ will decrypt to a plaintext that reflects the bit flips from the previous block: $p_2' = p_2 \oplus x$.
 
\begin{figure}[!tp]
  \centering
  \includegraphics[width=\linewidth]{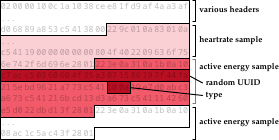}
  \caption{Anatomy of a vulnerable A-over-C message in plaintext. Each row corresponds to 16-byte cipher blocks.}
  \Description{An A-over-C plaintext annotated with the byte ranges containing a heart rate sample followed by two active energy samples. The UUID of the first active energy sample aligns with the 16-byte block size of AES. Two bytes containing type information are highlighted in the following block.}
  \label{fig:aovercMalleability}
\end{figure}

 We show that an attacker with access to A-over-C ciphertexts (Attacker~2) and partial knowledge of their plaintext content can use this property to change the type of transferred health samples, inserting forged values into the health database. This exploit relies on the 16-byte \glspl{uuid} present in transferred health samples: When this \gls{uuid} aligns with the blocks of the block cipher, we can flip bits in the ciphertext block containing it. The modified block will then decrypt to random bytes which, crucially, still form a valid random \gls{uuid}. We can then control the type byte of the health sample located in the following block---for example changing a heart rate into a step count. An illustrated example of this attack, performed on a real-world A-over-C health sync message, is shown in \autoref{fig:aovercMalleability}. Concretely, using \gls{cbc} malleability to flip the last four bits of the type field will make the receiver interpret the modified \textit{active energy} sample (0x0a) as a \textit{heartrate} sample (0x05), inserting the forged heart rate measurement into the database.

Notably, this behavior appears untouched by recent changes in iMessage cryptography~\cite{iMessagePQ}: Comparing our findings based on iOS 14.8 to the recent iOS 17.5.1, we find that A-over-C cryptography remains unchanged. Fortunately, the fix for this issue is simple: Apple could use the MessageProtection framework to encrypt the entire message rather than just encapsulate the key material, or use an authenticated encryption algorithm for the payload encryption.

\subsection{Health Data Deletion \& Cycle Tracking}

While not directly related to the Apple Watch, we also notice how health samples, most prominently symptoms logged in the cycle tracking app on either watch or phone, are---or rather are not---deleted. Some thought went into the security of this: Samples are not deleted using a basic \gls{sql} delete query, which might leave artifacts in the database file, but are manually marked as deleted and have most fields overwritten with null values. This removes, for example, the actual measurement contained in a heart rate sample. What stays, however, is the type of the sample and its deletion time. 

In the context of the cycle tracking app, this means that attackers can discover the number of deleted entries, their type, as well as the time they were deleted. While the database gives no information about the order or timing of these entries, this may still be troubling for users who have their devices seized by law enforcement as we briefly discussed in \autoref{sec:background:healthdata}: A set of symptoms associated with pregnancy that were deleted just before a device was seized might certainly look incriminating when being charged with an unlawful abortion. With no particular reason to keep deleted symptoms in the database, we do not see why these entries could not be overwritten and deleted entirely, thus avoiding this issue.

\section{Lessons for Secure Smartwatches}

\review{We add a section condensing what we learned from our analysis of the Apple Watch into general takeaways for deploying secure and private wearable devices. This also includes the importance of mechanisms for key rotation and tweakable crypto parameters, as rightly pointed out by reviewer C.}

At its core, the Apple Watch's security lies in \textbf{defense in depth} and strong, \textbf{standard protocols}. The watch does not have to rely on the varied security guarantees of lower-level transports, and the overall architecture remains largely secure even in the presence of the issues we discovered. Conversely, the security weaknesses we found stemmed from \textbf{proprietary protocol extensions} and \textbf{nonstandard cryptography}.

Based on these observations, we argue that a \gls{vpn} tunnel using well-understood security protocols can and should form the basis for future smartwatches. This tunnel can serve as a strong first protection layer that isolates higher-level logic from the lower-level transports and shields it from most network attackers. Any deviations from such standards, however, should be very carefully considered as they might have non-obvious security implications. Especially in cryptographic protocols, even small changes can have dire consequences. If non-standard cryptography cannot be avoided, it should therefore provide mechanisms for future flexibility: To keep systems secure in the long term, it is important to be able to change algorithms, update parameters, or rotate keys. \gls{ikev2}, for example, provides this flexibility through algorithm negotiation. Apple makes use of this, tweaking the preferred cipher suites from version to version. In the custom A-over-C protocol, on the other hand, parameters are hard-coded, and any change would break compatibility with prior versions---meaning that Apple is stuck with weak, unauthenticated cryptography for the foreseeable future.

\subsection{Other Smartwatch Architectures}
\label{sec:security:otherArchitectures}

\begin{figure}[t]
  \centering
  \includegraphics[width=\linewidth]{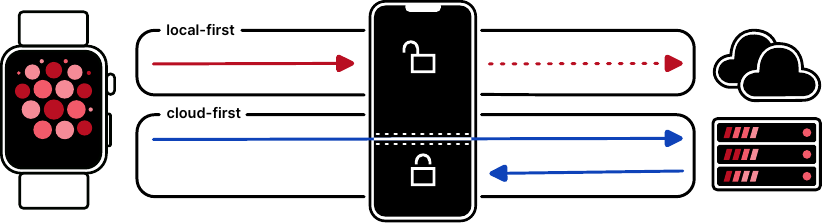}
  \caption{Local-first (top) versus cloud-first (bottom) smartwatch architectures.}
  \Description{In a local-first architecture, a smartwatch sends messages to the connected phone directly. The phone then optionally sends some messages to a cloud server. In a cloud-first architecture, the watch communicates directly with the server through an (encrypted) tunnel in which the phone acts as a proxy. The server then forwards some data back to the phone.}
  \label{fig:communicationStyles}
  \vspace{-1em} 
\end{figure}

When looking at wearable devices, we can distinguish two fundamental architectures that differ significantly in their privacy and security properties (see \autoref{fig:communicationStyles}): \textit{Local-first} devices communicate directly with a paired phone, and only involve remote servers for features that require explicit cloud synchronization. \textit{Cloud-first} devices only meaningfully communicate with a remote server, with the paired phone acting as a proxy that forwards messages to the cloud without the ability to perform any local processing.

The Apple Watch is a local-first device: Messages are always handled locally on the connected phone, and in most cases, no external server is involved at all. For messages that require a cloud service---such as Internet sharing or instant messaging---the phone forwards relevant payloads to a destination server. Notably, data is \textbf{end-to-end encrypted} between the watch and the phone.

Watches from Google-owned competitor Fitbit follow a cloud-first approach: Their devices establish an encrypted connection to Fitbit's servers using a factory-set key and a custom protocol~\cite{seemooFitbitVuln}. We confirm that the same architecture is present in the latest Google Pixel Watch. The phone still receives and forwards messages, but is not able to see their contents. Collected data is sent to the server using this encrypted connection before the server returns aggregated information to the phone.

While cloud-first devices offer better protection against local attackers by using pre-shared keys embedded in the hardware and communicating directly with a trusted server, this same server is also the main downside: A cloud-first scheme inherently places trust in a third party operating the server and cannot provide true end-to-end encryption between the user's devices. The third party will typically have access to plaintext user health data and extensive connection metadata. We therefore argue that a local-first model is preferable for user privacy and security as it facilitates end-to-end encryption, reduces metadata, and remains functional while offline.

\section{Related Work}

To the best of our knowledge, this is the first work to publicly reverse-engineer and document the communication protocols used by the Apple Watch, and the first demonstration of meaningful interoperability between the Apple Watch and an Android phone.

There is, however, prior work investigating the communication mechanisms of other fitness trackers, including devices from Fitbit~\cite{seemooFitbit, seemooFitbitVuln} and Xiaomi~\cite{breakmi}. Classen et al.~\cite{seemooFitbitVuln} in particular provide a full reimplementation of the Fitbit protocol stack alongside a custom app~\cite{seemooFitbitApp} that can receive fitness data from the tracker, similar to the custom app we develop to interact with the Apple Watch.

Early versions of Android Wear smartwatches were analyzed for security, especially with regard to physical access attacks~\cite{androidWearCaseStudy}. Similar scenarios of data extraction given physical access have been studied in the context of digital forensics, including case studies of smartwatches from Samsung and LG~\cite{forensicsSamsungLG} as well as Fitbit and Garmin~\cite{forensicsFitbitGarmin}. Health data collected by the Apple Watch has also been subject to forensics research; however, only synchronized data on the paired iPhone was considered~\cite{forensicsHealthdb}.

Some of the lower-level protocols for phone-to-watch communication are also used with other Apple devices and have been analyzed or described in these contexts. The \textit{Magnet} protocol and its message structures have been described by Heinze et~al.~\cite{toothpickerPaper}. The same paper also identifies the \textit{terminus} daemon as closely related to Apple Watch communication. The \textit{CLink} protocol, which we observe during early connection establishment over Bluetooth, also shows up in the work of Stute et~al.~\cite{alexContinuity} as \textit{Pair-Verify} and appears to be used with many Apple peripherals. Neither protocol, however, is of significant relevance for high-level watch communication.

Outside of academia, the Apple Watch has long attracted the attention of the jailbreaking community: Despite the lack of jailbreaks for modern watchOS versions, there is a variety of \textit{tweaks} for modifying the watch's behavior~\cite{legizmo, nanofi, winterMode, watchmutemirror}. As it is not currently possible to execute custom code not signed by Apple on modern watches,\footnote{The only publicly available watchOS jailbreak~\cite{watchOSjailbreak} targets the Apple Watch Series~3 running watchOS 4.1 and is, therefore, about seven years behind modern watch models.} these tweaks operate entirely on the paired (jailbroken) iPhone. Most of these tweaks are only cosmetic modifications to the notification logic on the iPhone, such as the \textit{WinterMode} tweak that makes notifications alert the user on both the watch and the phone, as opposed to the watch only~\cite{winterMode}. Some tweaks, however, hook into and modify the communication between watch and phone. \textit{WatchMuteMirror}~\cite{watchmutemirror}, for example, silences the iPhone when the user silences the watch, presumably listening for messages used to backup the watch's state to the phone to do so. 

The most advanced of these tweaks is \textit{Legizmo}~\cite{legizmoHomepage, legizmo}, which dramatically expands the version compatibility between iOS and watchOS beyond what is supported by Apple. This allows users to pair phones running older iOS versions with newer watchOS versions and vice versa. Legizmo developer \textit{lunotech11} told us that the tweak is based on extensive reverse engineering of the Apple Watch's communication protocols, bridging gaps and translating between different versions where necessary. These compatibility efforts go so far that Legizmo patches older apps to backport features only added in newer iOS versions, such as the advanced sleep tracking introduced in watchOS 9 and iOS 16. The tweak remains closed-source and we did not receive access for this work.

\section{Conclusion}

Our work on WatchWitch shows that true interoperability between the Apple Watch and third-party Android devices is feasible despite prior contrary claims: We have reimplemented several essential smartwatch features on Android, including push notifications, Internet sharing, and health data synchronization. We have also shown that we can achieve this level of interoperability while maintaining security---employing the same cryptographic protocols and storing keys and data with comparable hardware-backed security.

Going beyond interoperability, we have seen how opening the Apple Watch ecosystem to open-source implementations can benefit users by offering better privacy, more complete access to data, and even entirely new features such as a fine-grained firewall.

Our research makes the security and privacy properties of the Apple Watch visible and presents a way towards autonomy and independence, allowing users to use their devices on their own terms and beyond the manufacturer's intentions. We look forward to seeing researchers, tinkerers, and manufacturers build upon our work---be it in terms of alternative software, hardware, or entirely new applications.

\pagebreak 
\begin{acks}
With thanks to Apple for working with us throughout the disclosure process in fixing the vulnerabilities we discovered.
We also thank lunotech11 for insights on some of the Apple Watch's peculiarities.
This work has been funded by the German Federal Ministry of Education and Research and the Hessian State Ministry for Higher Education, Research, and the Arts within their joint support of the National Research Center for Applied Cybersecurity ATHENE.
\end{acks}

\bibliographystyle{ACM-Reference-Format}
\bibliography{bibliography}


\newpage
\appendix
\section{Appendix}

\subsection{Magnet}
\label{appendix:magnet}

The basic packet structure of Magnet has already been described by Heinze et~al.~\cite{toothpickerPaper}---we take a closer look at the Magnet handling logic in the Bluetooth daemon and identify the supported message types, shown in Table~\ref{tab:magnet:opcodes}. References to the debug string \texttt{"Received \%d remote services from the remote master \%p !"} may be used to find the main Magnet handling function in our iOS version (14.8) and likely many other versions as well.

\begin{table}[h]
\centering
\caption{Magnet opcodes and message types.}
\begin{tabular}{|l|l|}
\hline
\textbf{opcode} & \textbf{meaning}            \\ \hline
0x01            & remote services             \\ \hline
0x02            & remote services response    \\ \hline
0x03            & create channel for service  \\ \hline
0x04            & accept channel for service  \\ \hline
0x05            & service added               \\ \hline
0x06            & service removed             \\ \hline
0x07            & service removed acknowledge \\ \hline
0x08            & error response              \\ \hline
0x09            & version info                \\ \hline
0x70            & send time sync correction   \\ \hline
0x71            & time data                   \\ \hline
0x72            & time data                   \\ \hline
0x90            & DID info                    \\ \hline
0x91            & CL data                     \\ \hline
\end{tabular}
\label{tab:magnet:opcodes}
\end{table}

\subsection{NRLP}
\label{appendix:nrlp}

\gls{nrlp} is handled in the terminus daemon. As of iOS~14.8, the main parsing is performed in the \texttt{NRLink\b Bluetooth:\b handle\b ReadData} function. \gls{nrlp} packets may be fragmented across several \gls{l2cap} frames. Every \gls{l2cap} frame also contains a \textit{sequence number} and \textit{packets received} byte before the actual \gls{nrlp} data. As these bytes are also present for non-\gls{nrlp} traffic including \textit{CLink} and \textit{BT.TS}, we do not consider them to be part of \gls{nrlp}.

We list the supported payload types of \gls{nrlp} in \autoref{tab:nrlp:types}. Of these, we only see \gls{esp} and \gls{ikev2} related types in active use, with occasional Encapsulated6LoWPAN packets appearing as well. The echo service replies to \textit{ping} messages starting with the byte 0x01 with an identical \textit{pong} message starting with 0x02, but does not appear to be used.

\begin{figure}[h]
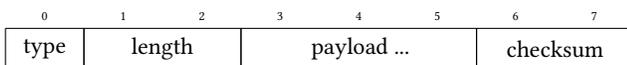

    \centering
    \begin{bytefield}[bitwidth=3.3em]{8}
        \bitheader{0-7} \\
        \bitbox{1}{type} & \bitbox{2}{length} & \bitbox{3}{payload ...} & \bitbox{2}{checksum}
    \end{bytefield}
    \caption{\gls{nrlp} message byte format.}
    \label{bytes:nrlp:structure}
    \Description{Byte composition of an \gls{nrlp} message, starting with a one-byte type field followed by a two-byte length field and a variable length payload. The message is concluded by a two-byte checksum.}
\end{figure}

The calculation of the checksum present at the end of each \gls{nrlp} packet (see \autoref{bytes:nrlp:structure}) differs between message types. For packet types below 0x64, the checksum is a common Internet Checksum as described in RFC~1071~\cite{rfcInternetChecksum}. For other messages, the checksum only covers the type and length header fields. The high and low byte of the checksum is calculated in this case as follows:

\begin{align*}
checksum_{high} &= length_{high} \oplus (type \ShiftRight 4) \\
checksum_{low} &= length_{low} \oplus (type \ShiftLeft 4)
\end{align*}

\begin{table}[h]
\centering
\caption{NRLP message types. ECT0 refers to the Explicit Congestion Notification flag of the Internet Protocol.}
\begin{tabular}{|l|l|}
\hline
\textbf{Type}   & \textbf{Meaning}    \\ \hline
0x00            & Pad0 / noop         \\ \hline
0x01            & PadN / noop         \\ \hline
0x02            & UncompressedIP      \\ \hline
0x03            & Encapsulated6LoWPAN \\ \hline
0x04            & IKEv2               \\ \hline
0x05            & echo service        \\ \hline
0x64            & ESP                 \\ \hline
0x65            & ESP\_ECT0           \\ \hline
0x66            & TCP                 \\ \hline
0x67            & TCP\_ECT0           \\ \hline
0x68            & ESP\_ClassC         \\ \hline
0x69            & ESP\_ClassC\_ECT0   \\ \hline
\end{tabular}
\label{tab:nrlp:types}
\end{table}

\subsection{IKEv2 Custom Notify Payloads}
\label{appendix:ike}

As of iOS~14.8, the IKEv2 handshake with the Apple Watch is also handled by the terminus daemon. Handling of the private notify payloads in particular is performed in the \texttt{NRLinkBluetooth\b::\b handle\b NotifyCode:\b payload:} method.

\begin{figure}[h]
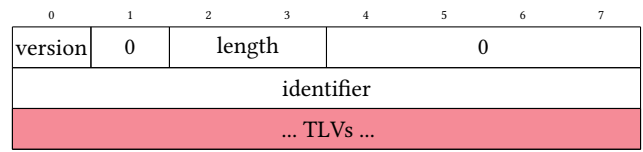

    \centering
    \begin{bytefield}[bitwidth=3.3em]{8}
        \bitheader{0-7} \\
        \bitbox{1}{version} & \bitbox{1}{0} & \bitbox{2}{length} & \bitbox{4}{0} \\
        \bitbox{8}{identifier} \\
        \bitbox[tlrb]{8}[bgcolor=witchlightred]{... TLVs ...} \\
    \end{bytefield}
    \caption{Structure of the \gls{ldm} notify payload. The only observed version is 2, the length field encodes the byte length of the TLVs only.}
    \Description{\glspl{ldm} start with a version byte followed by a constant zero byte, a two-byte length field and four more zero bytes. After that is an eight byte identifier followed by a variable length TLV-encoded payload section.}
    \label{bytes:ikeLinkDirectorMessage}
\end{figure}

Beyond the \glspl{ldm} used for \mbox{Wi-Fi} discovery, Apple uses a variety of other private notify payloads for signaling purposes. This includes the communication of version information as well as tunnel IP addresses and various configuration flags. An overview of all notify types is shown in~\autoref{tab:ike:notify}. The byte structure of \acrfull{ldm} payloads in shown in \autoref{bytes:ikeLinkDirectorMessage}. The \gls{ldm} TLVs (\autoref{tab:ike:ldm}) other than the previously mentioned Update \mbox{Wi-Fi} Address messages are related to management of the wireless links and appear to be only rarely used.

\begin{table}[!tp]
\centering
\caption{Private notify types used by Apple in their IKEv2 implementation. \textit{IAdr} short for \textit{InnerAddress}.}
\begin{tabular}{|l|l|p{3.8cm}|}
\hline
\textbf{ID} & \textbf{Name}               & \textbf{Comment}                                                      \\ \hline
48601              & Encrypted prelude           & Bluetooth only, echoes prelude sent at the start of a NRLP connection \\ \hline
48602              & Terminus version            & e.g. 0x00d, 0x00c                                                     \\ \hline
48603              & Device name                 & e.g. "iPhone", "Apple Watch"                                          \\ \hline
48604              & Build version               & e.g. "18H17", "18S830"                                                \\ \hline
50701              & ProxyNotify                 & IPv6 address and port of Shoes server on the phone                    \\ \hline
50702              & LinkDirectorMessage         & used for link state signaling and \mbox{Wi-Fi} discovery                      \\ \hline
50801              & IAdrInitiatorClassD & IPv6 tunnel address used by the watch for class D traffic  \\ \hline
50802              & IAdrResponderClassD & IPv6 tunnel address used by the phone for class D traffic  \\ \hline
50811              & IAdrInitiatorClassC & IPv6 tunnel address used by the watch for class C traffic  \\ \hline
50812              & IAdrResponderClassC & IPv6 tunnel address used by the phone for class C traffic  \\ \hline
51401              & Always-On \mbox{Wi-Fi}      & 1 byte boolean flag                             \\ \hline
51501              & IsAltAccountDevice          & 1 byte boolean flag                                 \\ \hline
\end{tabular}
\label{tab:ike:notify}
\end{table}

\begin{table}[h]
\centering
\caption{Types of Link Director Message \glspl{tlv}.}
\begin{tabular}{|l|l|p{4.05cm}|}
\hline
\textbf{\#}   & \textbf{Name}         & \textbf{Comment}                    \\ \hline
\cellcolor{witchlightred}1 & Hello                 & no payload, signals restart         \\ \hline
\cellcolor{witchlightred}2 & UpdateWiFiAddressIPv6 & 2 byte port followed by 16 byte IP  \\ \hline
\cellcolor{witchlightred}3 & UpdateWiFiAddressIPv4 & 2 byte port followed by 4 byte IP   \\ \hline
\cellcolor{witchlightred}4 & UpdateWiFiSignature   & variable length                     \\ \hline
\cellcolor{witchlightred}5 & PreferWiFi            & no payload                          \\ \hline
\cellcolor{witchlightred}6 & DeviceLinkState       & 1 byte, 1: Bluetooth, 2: \mbox{Wi-Fi}       \\ \hline
\cellcolor{witchlightred}7 & PreferWiFiAck         & 1 byte boolean                      \\ \hline
\cellcolor{witchlightred}8 & ForceWoW              & no payload, WoW is Wake-on-Wireless \\ \hline
\end{tabular}
\label{tab:ike:ldm}
\end{table}

\pagebreak

\subsection{Cryptographic Algorithms}
\label{appendix:crypto}

As explained in \autoref{sec:communication}, the watch uses different modes of encryption when communicating with iOS. This section, focuses on the cryptographic algorithms used and their security aspects.

\subsubsection{IKEv2/ESP}

We include a full list of cryptographic primitives supported in \gls{ikev2} for the watch models we tested in \autoref{tab:ike:watchAlgosIKE}. The algorithm identifiers match the constants defined by IANA for use in \gls{ikev2}~\cite{ianaIkeParams}. For the Series~5 watch, we can observe that the \gls{esp} encryption algorithms match the ones offered for \gls{ikev2}. However, on newer Apple Watch models we are unable to observe the provided algorithms due to a lack of a jailbreak for iOS~17 or watchOS~10 at the time of writing. We expect that the \gls{esp} algorithms also match the algorithms offered for \gls{ikev2} for these newer versions.

\begin{table}[h]
\centering
\caption{Cryptographic algorithms advertised by watch models for use in IKEv2, in order of preference.}
\begin{tabular}{|ll|}
\hline
\multicolumn{1}{|c|}{\cellcolor{gray!30}Series 5, watchOS 7.3.3}     & \multicolumn{1}{c|}{\cellcolor{gray!30}Series 9, watchOS 10.0.2} \\ \hline
\multicolumn{2}{|l|}{\textbf{Encryption}}                                      \\ \hline
\multicolumn{1}{|l|}{ChaCha20-Poly1305}           & AES-GCM-16 (256bit)        \\ \hline
\multicolumn{1}{|l|}{AES-GCM-16 (256bit)}         & ChaCha20-Poly1305          \\ \hline
\multicolumn{2}{|l|}{\textbf{Pseudo-Random Function}}                          \\ \hline
\multicolumn{1}{|l|}{HMAC-SHA2-512}               & HMAC-SHA2-512              \\ \hline
\multicolumn{1}{|l|}{HMAC-SHA2-256}               &                            \\ \hline
\multicolumn{2}{|l|}{\textbf{Diffie-Hellman Group}}                            \\ \hline
\multicolumn{1}{|l|}{Curve25519}                  & Curve448                   \\ \hline
\multicolumn{1}{|l|}{521-bit random ECP group}    & Curve25519                 \\ \hline
\multicolumn{1}{|l|}{8192-bit MODP Group}         &                            \\ \hline
\multicolumn{2}{|l|}{\textbf{Signature Hash Algorithm}}                        \\ \hline
\multicolumn{1}{|l|}{SHA2-256}                    & Identity                   \\ \hline
\multicolumn{1}{|l|}{Identity}                    & SHA2-256                   \\ \hline
\end{tabular}
\label{tab:ike:watchAlgosIKE}
\end{table}

\paragraph{Diffie-Hellmann Key Exchange}

The Apple Watch primarily uses \gls{DH} key exchanges based on \gls{ecc}, allowing for smaller key sizes with security levels comparable to larger modular exponentiation groups. 
When using \gls{ecc}, the use of an appropriate curve is paramount to its security. Apple decided to select only curves which result in 256-bit symmetric keys and all of these curves are standardized by the IETF~\cite{kojoMoreModularExponential2003,fuECPGroupsIKE2007,nirCurve25519Curve448Internet2016}. Curve448 and Curve25519 are the recommended curves to be used for security purposes~\cite{langleyEllipticCurvesSecurity2016} and are the only curves supported from watchOS~10 on.

\paragraph{Encryption}

To perform symmetric encryption in \gls{ikev2} and \gls{esp}, ChaCha20-Poly1305 or AES-GCM-16\footnote{\textit{16} denotes the number of octets used for the authentication tag.} with 256-bit keys are used. Both algorithms are \gls{AEAD} schemes, authenticating the encrypted message and optional additional plaintext data using an authentication tag. When decrypting the ciphertext, the algorithm checks if the authentication tag matches the expected value and throws an error if not. An adversary in a \gls{mitm} position modifying the ciphertext or authenticated data can be detected and the integrity of the message is protected.
The formal security of both ciphers has been proven~\cite{niwaGCMSecurityBounds2015,degabrieleSecurityChaCha20Poly1305MultiUser2021} and there exist no known attacks against them.

\subsubsection{A-over-C}

Apple uses \gls{ikev2} and \gls{esp} for general data transfer between the Apple Watch and the connected iPhone. However, for sensitive data they add a second layer of a custom encryption scheme called A-over-C (see ~\autoref{sec:aoverc}). 
A-over-C uses RSA-\gls{OAEP} with 1280-bit keys and AES-CTR (counter mode) with a 128-bit key to encapsulate an ephemeral 128-bit key, which is chosen randomly for every message. The encapsulated key is authenticated using an ECDSA signature with a 384-bit key.
The resulting encapsulation scheme provides confidentiality and authenticity, and has been formally analyzed in the context of iMessage, which uses an identical construction~\cite{imessageAttack}.\footnote{The malleability attacks from~\cite{imessageAttack} do not apply here as the entire encapsulated key is contained in the RSA-OAEP ciphertext, which is non-malleable.} While key sizes for AES and ECDSA match recommended values, the RSA keys fall well short of the recommended minimum of 2048~bit~\cite{nistKeysize}, making them vulnerable to potential brute-force attacks in the future. 

The payload itself is encrypted using AES-CBC (cipher block chaining mode). Like AES-CTR, AES-CBC provides confidentiality if used correctly but does not provide authentication as would be the case for \gls{AEAD} schemes such as AES-GCM. We detail this issue in \autoref{sec:security:forging_health_data}. AES-CBC as used by the Apple Watch is also vulnerable to \textit{padding oracle} attacks which could reveal message plaintext under the right circumstances~\cite{cbcPaddingOracle}.

\clearpage

\end{document}